\RequirePackage{fix-cm}
\documentclass[runningheads]{llncs}
\newcommand{\y}[1]{#1} 
\newcommand{\yy}[1]{} 
\newcommand{\yyy}[1]{} 
\newcommand{\hab}{\hspace*{23em}} 
\newcommand{\ab}{\scriptsize (a)\hab (b)} 

\newcommand{\cd}{\scriptsize (c)\hab (d)}

\usepackage{graphicx}
\RequirePackage{graphicx}
\usepackage{underscore} 
\usepackage{bm}
\usepackage{cite}
\usepackage{color}
\usepackage{pmboxdraw} 
\usepackage{url} 
\makeatletter
\AtBeginDocument{%
  \renewcommand{\doi}[1]{\url{https://doi.org/#1}}}
\makeatother
\definecolor{ggreen}{rgb}{0,.45,0}
\definecolor{gred}{rgb}{.8,0,0}
\definecolor{gblue}{rgb}{0,0,.7}
\definecolor{ForestGreen}{cmyk}{0.91,0,0.88,0.12} 
\definecolor{MyGray}{gray}{0.5}
\definecolor{ggreen}{rgb}{0,.47,0}
\newcommand{\aB}[1]{{\color{black}#1}}    

\usepackage{amsmath,amssymb,amsfonts,bm} 
\usepackage{floatfig,wrapfig,epsfig}
\usepackage{subfigure}
\usepackage{psboxit}
\usepackage{rotating}
\usepackage{curves}
\usepackage{ulem}
\usepackage[mathscr]{eucal}
\RequirePackage{psfrag}
\usepackage{soul}           
\usepackage{pmboxdraw} 

\newcommand{\chose}[2]{\genfrac{(}{)}{0pt}{}{#1}{#2}}

\makeatletter
\newcommand*\rel@kern[1]{\kern#1\dimexpr\macc@kerna}
\newcommand*\widebar[1]{%
  \begingroup
  \def\mathaccent##1##2{%
    \rel@kern{1.2}
    \overline{\rel@kern{-1.25}\macc@nucleus\rel@kern{-0.1}}
    \rel@kern{0.1}%
  }%
  \macc@depth\@ne
  \let\math@bgroup\@empty \let\math@egroup\macc@set@skewchar
  \mathsurround\z@ \frozen@everymath{\mathgroup\macc@group\relax}%
  \macc@set@skewchar\relax
  \let\mathaccentV\macc@nested@a
  \macc@nested@a\relax111{#1}%
  \endgroup
}
\makeatother

\newcommand{\eq}[2]{\begin{equation}\label{#1}#2\end{equation}}
\newcommand{\eqs}[1]{\begin{eqnarray*}#1\end{eqnarray*}}

\newcommand{\beq}[1]{\begin{equation}\label{#1}}
\newcommand{\eeq}{\end{equation}}
\newcommand{\beg}{\begin{gather*}}
\newcommand{\eeg}{\end{gather*}}
\newcommand{\beqq}{\begin{eqnarray}}
\newcommand{\eeqq}{\end{eqnarray}}
\newcommand{\x}[1]{} 
\newcommand{\E}{\mathbb E} 
\newcommand{\cdc}{,\ldots,}

\newcommand{\seRed}{\left\lfloor\frac n2\right\rfloor}

\def\wE{4.0} 
\def\hEh{5.3}
\def\hEEh{4.95}
\def\hFh{4.1}

\def\wF{4.5} 

\def\wFH{6.1} 

\def\xyz{\hspace*{.07em}}
\def\xy{\hspace*{.12em}}

\def\xz{\hspace*{-.07em}}

\def\E{{\mathbb E}}
\def\lL{\ell}
\def\Es{\E(d^\mathrm{\xy str}_{\xi})}  %

\def\Up#1{\vspace*{-#1em}}

\def\Eg#1{\E(d^\mathrm{\xy gr#1})} 

\def\Ei{\E(d^\mathrm{\xy ind})}  

\def\fOl{f^{\xyz0,1\xz}}
\def\Nmusi{N^{\mu,\sigma}}

\def\NOl{N^{0,1}}
\def\NOlt{N^{0,12}}
\def\sig{\sigma_{\!g}}

\begin{document}\normalem
\title{Evolution of Society Caused by Collective and Individual Decisions\x{: A~ViSE Model Study} 
}
\subtitle{A~ViSE Model Study}
\titlerunning{Evolution of Society Caused by Collective and Individual Decisions 
}

\author{Pavel Chebotarev\x{\inst{1}}\orcidID{0000-0001-8232-847X}
}

\institute{Technion--Israel Institute of Technology, Haifa, 3200003 Israel \email{pavel4e@technion.ac.il; pavel4e@gmail.com} 
}

\maketitle

\begin{abstract}
\aB{Decision-making societies may vary in their level of cooperation and degree of conservatism, both of which influence their overall performance. Moreover, these factors are not fixed---they can change based on the decisions agents in the society make in their interests. But can these changes lead to cyclical patterns in societal evolution?
To explore this question, we use the ViSE (Voting in Stochastic Environment) model. In this framework, the level of cooperation can be measured by group size, while the degree of conservatism is determined by the voting threshold. Agents can adopt either individualistic or group-oriented strategies when voting on stochastically generated external proposals.
For Gaussian proposal generators, the expected capital gain (ECG)---a measure of agents' performance---can be expressed in standard mathematical functions. Our findings show that in neutral environments, societal evolution with open or democratic groups can follow cyclic patterns. We also find that highly conservative societies or conservative societies with low levels of cooperation can evolve into liberal (less conservative than majoritarian) societies and that mafia groups never let their members go when they want to.}
\x{}
\end{abstract}


\section{Introduction}

\x{A model}Settings in which agents\x{participants} vote for projects formulated in terms of their\x{ [positive or negative]} capital gains were analyzed by A.V.~Malishevskii in the early 1970s; 
some of these results were presented in \cite{Mirkin79} and \cite{Aizerman81e}. In Malishevskii's model, voting is easily manipulated by agenda-setters\x{the organizers}. Similar results have been obtained in frameworks in which participants have ideal points in a multidimensional program space \cite{McKelvey76}\x{; see also \cite{Aizerman81e}}. Spatial models of voting have been studied extensively in \cite{EnelowiHinich84}. \x{The}Problems of the relationship between egoism, altruism, collectivism, and rationality were considered in 
\cite{Margolis84Selfishness,Levine98Modeling,Lindenberg01}
and voting as a method for making decisions about redistribution of social benefits by means of taxation and social programs was discussed
\aB{in
\cite{Romer75IndWelf,Roberts77Voting,Kranich01AltruIncome,Galasso02EconSecu} among others. 
\x{We are interested in the average increments of the payoff vectors rather than in the payoff vectors at the end of the play.}

Two important factors that influence the performance of decision-making societies are the level of cooperation and the degree of conservatism.
Therefore, agents may want\x{wish.5} to change them in their own interests. What could be the trajectories of such successive reforms? In this paper, we study this question using the ViSE model. Within its framework\x{this model}, the level of cooperation and the degree of conservatism can be measured by the group size and the voting threshold, respectively. We find that the\x{ societal.3} evolution of a society in terms of these parameters can have cyclical and some other specific patterns.
}

Recent work related to our study is discussed in~\cite{AfoChe25}. 

\subsection{The ViSE Model} 

The main assumptions of the ViSE model (see \cite{MaksChe20,AfoChe25}) are as follows. 
A {\em society\/} consists of $n$ {\em agents\/} (also called {\em voters}). Each agent is characterized
by the current value of their {\em capital\/} (\aB{a real number}; debt, if negative), which can sometimes be interpreted as utility.
In each step\x{ $m = 1\cdc M$}, some proposal is put to the vote, and the agents vote, guided by their\x{ individual voting mechanisms} strategies, for or against it. Within the ViSE model, a strategy is understood as an algorithm following which an agent uses the available information to decide whether to vote for or against the proposal under consideration. 
Each {\em proposal\/} is a vector of algebraic capital gains of all agents. Proposals approved using the\x{an} established voting rule are implemented.

\x{In the ViSE model, e Each}Proposals are generated stochastically: their components are realizations of random variables. \x{In this paper, }We consider the\x{ simple} case where the components\x{ of a/the proposal} are independent and identically distributed with a given mean~$\mu$\x{ and standard deviation~$\sigma$}. The corresponding scalar random variable $\xi$ is called the {\em gain generator}. The proposals put to the vote can be called {\em stochastic environment proposals}.
The environment is {\em favorable\/} if $\mu > 0,$ {\em neutral\/} if $\mu = 0,$ and {\em unfavorable\/} ({\em hostile\/}) if $\mu < 0.$

The dynamics of agents' capital in various environments can be analyzed to compare voting strategies and social decision rules in order to select the optimal ones in terms of maximizing appropriate criteria.
Since the process in the presented version of the model is stationary, it can be described by one-step characteristics. An important one is the\x{ [negative or zero or positive]} mathematical expectation of one-step capital gain for an agent with a given strategy after\x{ rejecting or accepting and} implementing the collective decision regarding the submitted proposal using\x{through} the established\x{adopted} voting rule. This value is called the {\em expected capital gain $($ECG\/$)$ of an agent\/} (this abbreviation was used in \cite{Weron98,PavlovaRigobon10,TsChLo20Cambr}, etc.\x{; not to be confused with electrocardiogram})
and denoted by $\Es$, where \aB{abbreviations of agent strategies can be substituted for $\mathrm{str}$. For the societies considered in this paper, the ECG is determined by Corollary~\ref{c:ECG} formulated\x{ presented} in Section~\ref{s:corol}}. 

The properties of the environment influence the relationship between the current and future states of society.
In this regard, the model is relevant to situations where the issue\x{ question} is one of comparing the {\em status quo\/} with reform, rather than choosing\x{ between several projects or} among several candidates. 

This paper examines the decisions made with\x{ simple majority voting and} two agent strategies. 
An {\em individualist\/} (1-{\em agent}) supports a proposal if and only if it increases their own capital.
The voting strategy of a {\em group member\/} is to support a proposal if and only if this proposal
increases the group's total capital.
We consider a\x{n inhomogeneous} society in which $\lL$ participants are individualists 
and $g$ agents form a group, so that
$g+\lL=n$
is the size of the society. 
We denote this society $I|G$ and call it a {\em single-group\/} \aB{or} {\em two-component\/} society.
A society is {\em individualistic\/} ( respectively, {\em clique-like\/}) if $\lL=n$ (resp., $g=n$).

The votes are aggregated using the $t$-majority rules: a proposal is accepted if and only if more than $t$\ ($t\in[-1,n]$ is a parameter) participants vote for it.
Thus, not only a {\em $[$simple\/$]$ majority} ($t=\seRed$) is considered, but also {\em qualified majorities\/} ($t\ge\seRed+1$) often used to make constitutional or other important decisions, and {\em initiative minorities\/} with $t<\seRed,$ used for such decisions as including issues on the agenda, forming new \aB{parliamentary} groups, submitting requests, initiating referendums, etc.

\aB{Societies will be called {\em conservative} (resp., {\em majoritarian}, {\em liberal\/})} if $t\ge\seRed+1$ (resp., $\seRed\le t<\seRed+1$, $t<\seRed$); $t$ measures the degree of conservatism.
The parameter $g$ expresses the degree of {\em cooperation\/} in society.
\aB{A relative measure of cooperation is~$g/n$.}
Thus, we \aB{classify} societies in terms of \x{liberalism}conservatism and cooperation.

One of the questions under study is: How does the effectiveness of the $t$-majority voting and the effectiveness of the individualistic and group strategies depend on the parameters $t$, $n$, $\lL$, and the favorability of the environment?

For a discussion of the relation of the ViSE model to reality, 
we refer to \cite{MaksChe20}. Connections with other models are indicated in~\cite{AfoChe25}.
\aB{Early results on the ViSE model (studying the ECG in two-component societies with different group types and voting thresholds) were obtained in~\cite{Che06ARC}. For more results on two-component societies, see~\cite{CheMax21}. The ViSE model implements \cite{CheLog10ARC} the well-known ``small party bias'': in the presence of two large parties or coalitions, neither of which has a majority, a small party forming a majority with any of them gains an advantage, which can be comparable to a dictatorship when large coalitions take opposite positions. The model also reveals one of the stability sources of the bipartisan system with almost equal parties. The ViSE model implements \cite{CheLog10ARC,CheMax21} the ``snowball'' scenario of cooperation: since belonging to a group is usually more beneficial than protecting one's own interests individually, participants join the group, and as it grows, group egoism becomes closer to altruism. In \cite{CheMal18opt,Malyshev21optimal}, the pit of losses paradox was discussed and the problem of the optimal voting threshold has been solved. A ``responsible elite'' can help overcome the pit of losses paradox~\cite{TsChLo20Cambr,TsoChe24E}. Another possible solution is based on taxes~\cite{Afonkin21tax}. In \cite{CheTsoLog18,MaksChe20}, it was found that the tail heaviness of the distribution of random variables generating proposals affects the effectiveness of agent's strategies.
}

\subsection{Research Framework} 

It is not uncommon for people to unite into teams or groups and defend group interests rather than individual ones. In fact, this is the background to\x{of6} almost every historical event.
Group interest may differ greatly from individual interest, so the question of whether this strategy is rational is non-trivial. This issue is studied within various\x{ a plethora of} models including game-theoretic ones. The ViSE model is useful\x{ very convenient} for such studies, as it provides convenient means for varying essential\x{important} parameters, such as the favorability of the environment, the structure of the society, the  social decision rule\x{voting threshold}, etc. In this paper, we study how the performance of individualistic and group agents depends on the voting threshold and the structure of\x{ last two parameters} a two-component\x{ and three-component} society.

Furthermore, the teams/groups can appear, increase, decrease, merge, split, and disappear. Individualists may join groups\x{; conversely}, while group members may become individualists or form other groups when this is beneficial to them\x{ [\hl{repeated in Section}~\ref{s:2compon}]}.
Finally, changing the voting threshold $t$ can be beneficial for some categories of agents and disadvantageous for others. Decisions on changing $t$ can be made by majority voting or voting with the current threshold.
In this paper, we study the consequences and benefits of such transitions.
The decisions mentioned determine the dynamics of society.

\x{On the other hand Furthermore, }
This dynamics essentially depends on the \x{rights}type of the group. An {\em open\/} group accepts\x{admits} all individualists willing to join it and lets its members go \x{if}whenever they wish. A {\em democratic\/} group always allows {\em leaving\/} it and allows {\em entering\/} it iff it is beneficial to the current members of the group. A {\em mafia}-group \aB{only} allows \x{transitions}\aB{changes to it} that are beneficial for its current\x{present} members.\x{It turns out The study will establish that a mafia group {\em never\/} lets go its members that wish to go.}
The study found that a mafia group {\em never\/} lets go of its members who want to leave.

\aB{The purpose of this paper is to explore the possible evolution of such societies.  
In particular, can the evolution patterns be cyclical?  We show that this is the case. 
For example, a society with a small group and low voting threshold will approve of raising its threshold, then increase the size of the group while gradually\x{ slowly} lowering the threshold, then support a decrease in the size of the group, which will return the society to its original position.

The study is based on Corollary~\ref{c:ECG}\x{, an general analytical result} presented in Section~\ref{s:corol}. We consider societies consisting of $n=25$ agents. It was found that for larger $n$, the main regions of the diagrams remain the same, while additional details may appear at the boundaries of the regions. For smaller $n,$ some details disappear. We consider the Gaussian proposal generator as the simplest one. The difference in results for other continuous symmetric distributions is not fundamental; for more information on the dependence on distributions, see~\cite{CheTsoLog18,MaksChe20}.
} 

\section{Expected Capital Gains in a Single-Group Society}\label{s:corol}

\begin{corollary}[\!\!\xz\cite{CheMax21}]\label{c:ECG}
For a society of $n$ agents with a group size $g$ and a voting threshold $t,$ under a Gaussian proposal generator $N^{\mu,\sigma},$ the ECGs of individualists and group members are expressed as follows\/$:$
\eqs{
\Ei
&=&\left(r(P_{t-g-1,\lL-1}-P_{t-g,\lL})+\mu P_{t-g,\lL}\right)\NOl(\mu_g)\\
&+&\left(r(P_{t  -1,\lL-1}-P_{t,  \lL})+\mu P_{t,  \lL}\right)\NOl(-\mu_g);
}

\Up{1.4}$$
\Eg{}
=      \left( P_{t-g,\lL}-P_{t,\lL}\right)\left(\mu \NOl(\mu_g)+\sig\xy\fOl(\mu_g)\right)+\mu P_{t,  \lL},
$$
where\\[-12pt]
\eq{eq:Ppkm}{
P_{k,m}=
\begin{cases}
                                                                                     1, & \phantom{-1\le }\;k<-1\\
\sum_{i=k+1}^m\chose m{i}p^iq^{m-i} = 1-\mathrm{Bi}(k\,|\,m,p)\x{=\mathrm B(p\,|\,k+1,m-k)
                                                             \mathrm B_{k+1,m-k}(p)},   &          -1\le  k<m\\
                                                                                     0, & \phantom{-1\le }\;k\ge m,
\end{cases}
}\\[-9pt]
 $\mathrm{Bi}(\cdot\,|\,m,p)$      is the binomial CDF with $m$ trials and success probability $p,$
$p=1-q,$
$q=\x{\Nmusi(0)}\NOl\!\left(\frac{\mu}{\sigma}\right),$
$r=\frac{\sigma\tilde f}{q},$\x{ auxiliary constant}
$\tilde f=\fOl\!\left(\frac{\mu}{\sigma}\right),$\x{ auxiliary density constant}
$\mu_g=\frac\mu{\sig},$\x{``adjusted mean'' for the group }
$\sig=\frac\sigma{\sqrt{g}},$\x{ standard deviation of the average for the group}
$\fOl(\cdot)$ and $\NOl(\cdot)$ are the standard \aB{normal} PDF and CDF$,$ respectively$.$
\end{corollary}

\Up{1}
\aB{
\begin{remark}
Using the connection between the binomial and Beta distributions we have in \eqref{eq:Ppkm}: $1-\mathrm{Bi}(k\,|\,m,p)=\mathrm B_{k+1,m-k}(p),$ where
$\mathrm B_{\alpha,\beta}(\cdot)$ is the CDF of the Beta distribution with shape parameters $\alpha>0$ and $\beta>0.$
\end{remark}
}

\begin{figure}[t!]\begin{center} 
\y{%
\includegraphics[height=\wE cm]{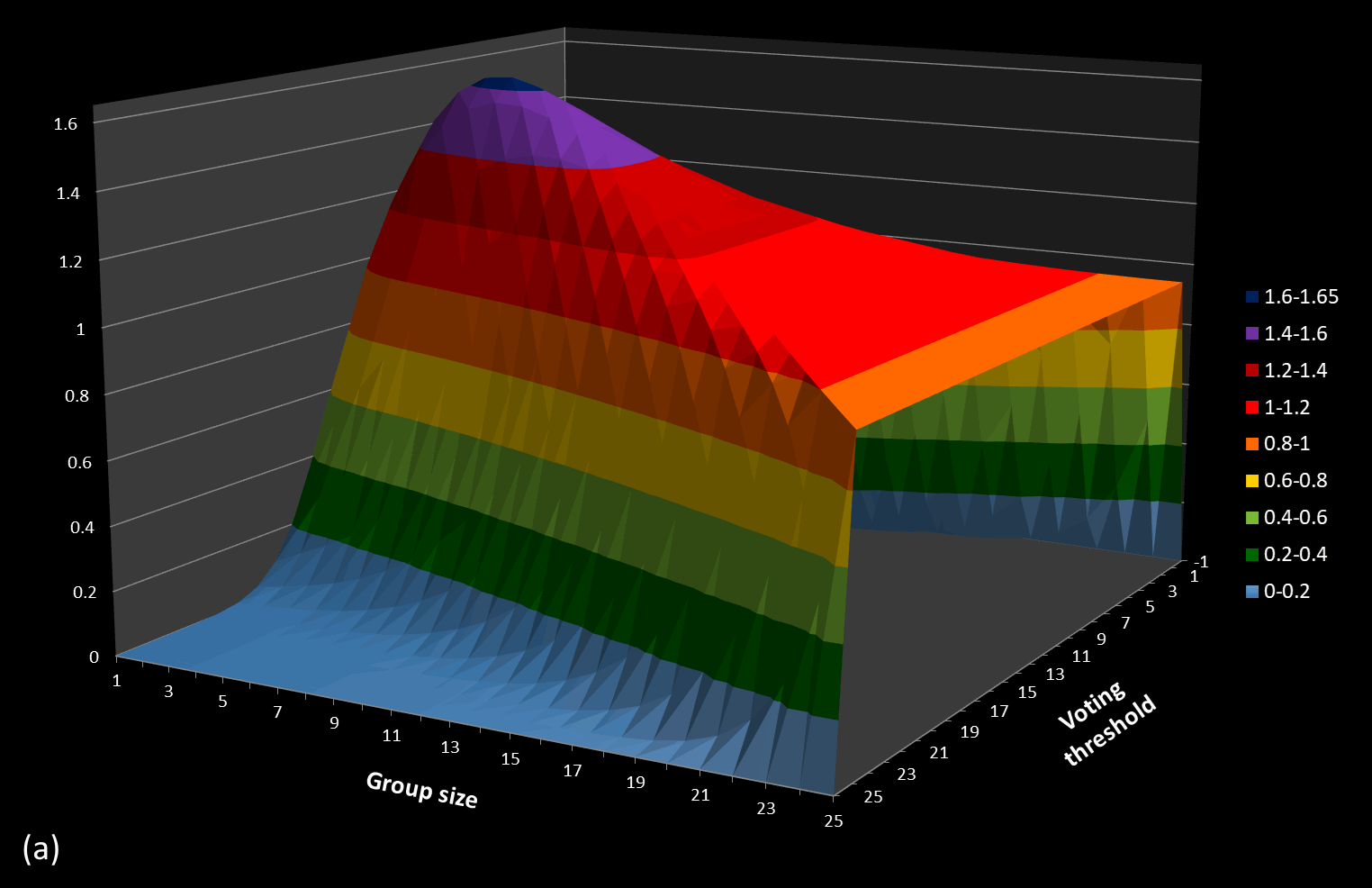}\xy\includegraphics[height=\wE cm]{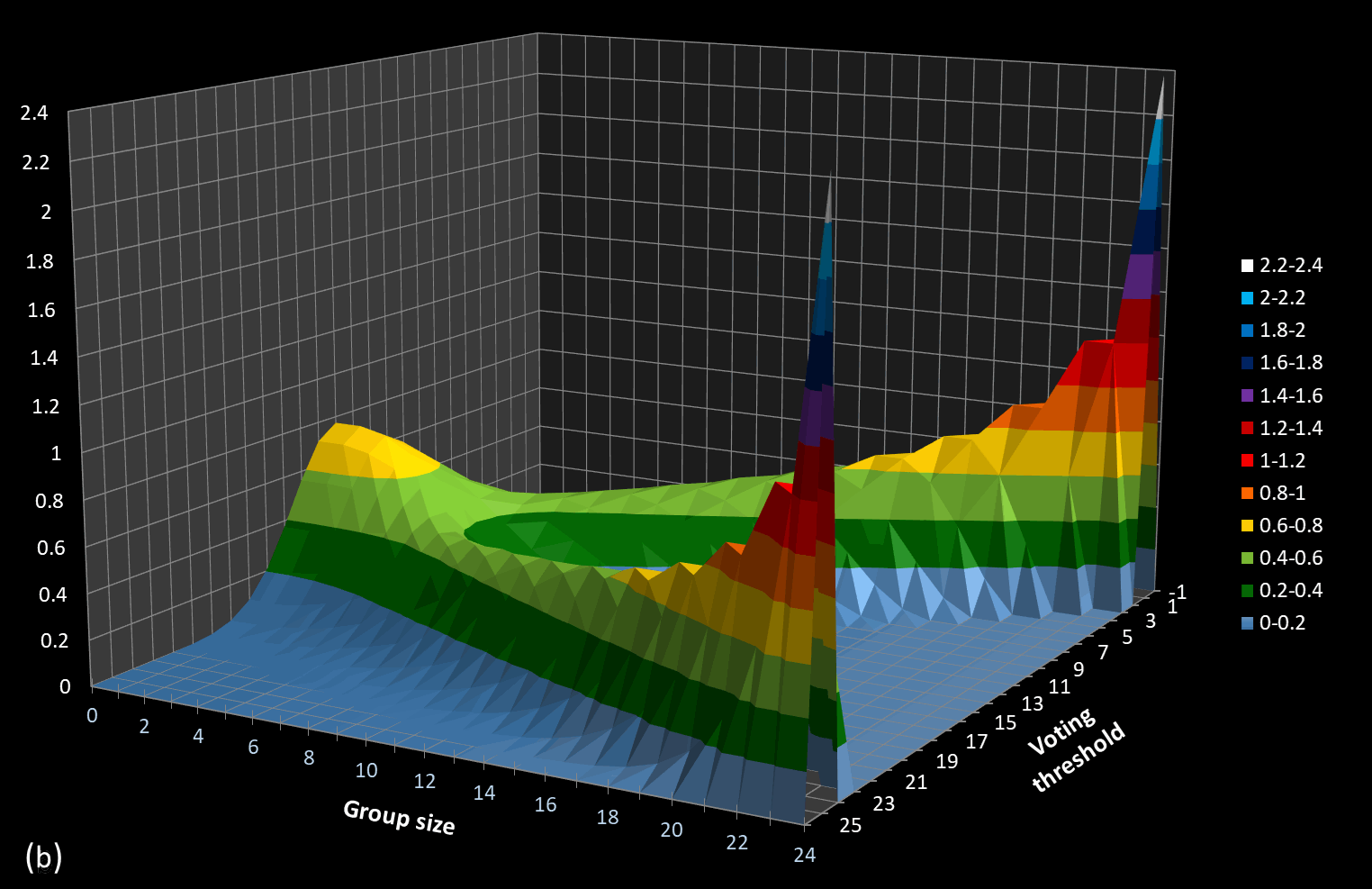}\\{\hspace*{.03em}}%
\includegraphics[height=\wE cm]{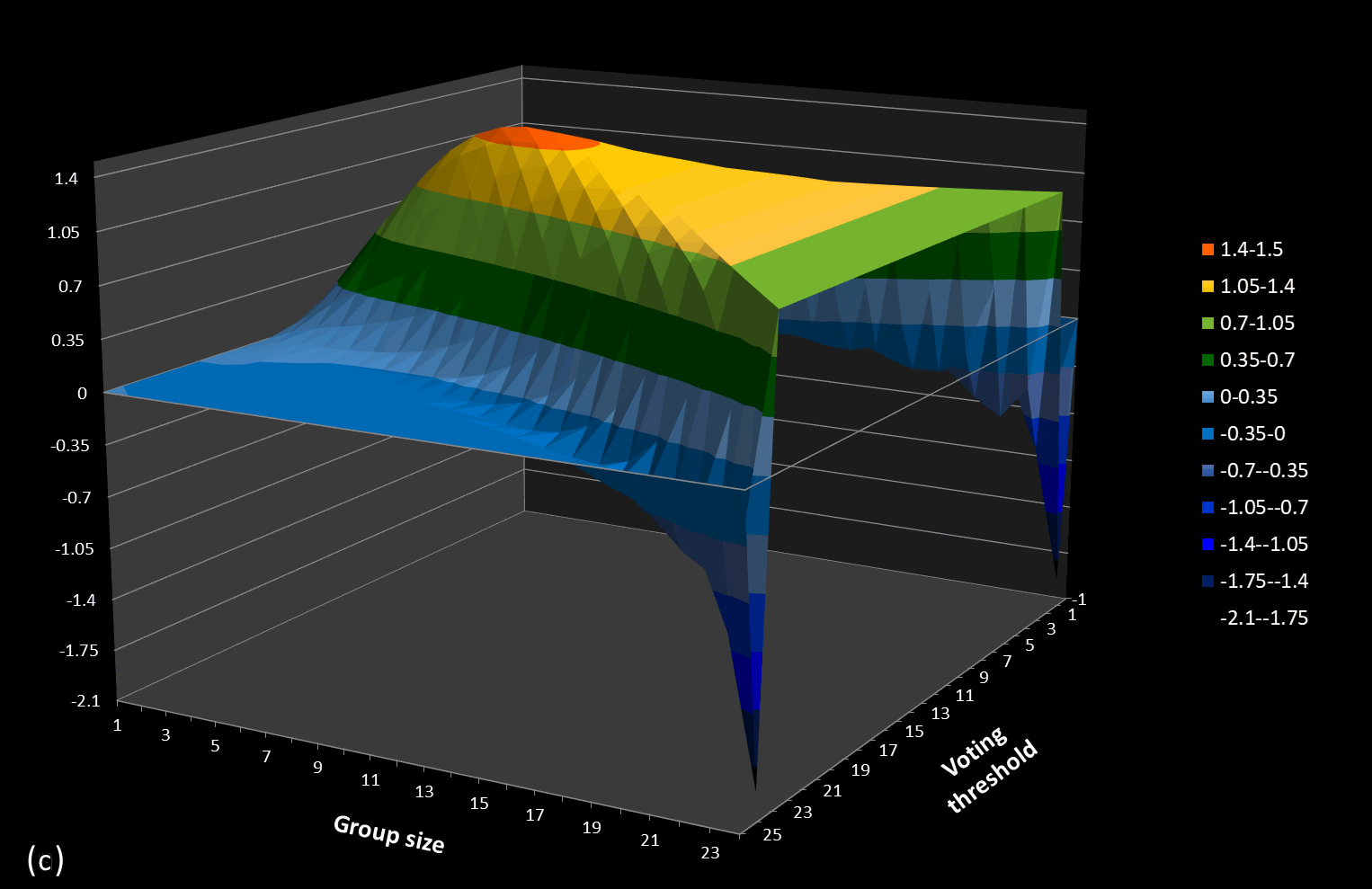}\xy\includegraphics[height=\wE cm]{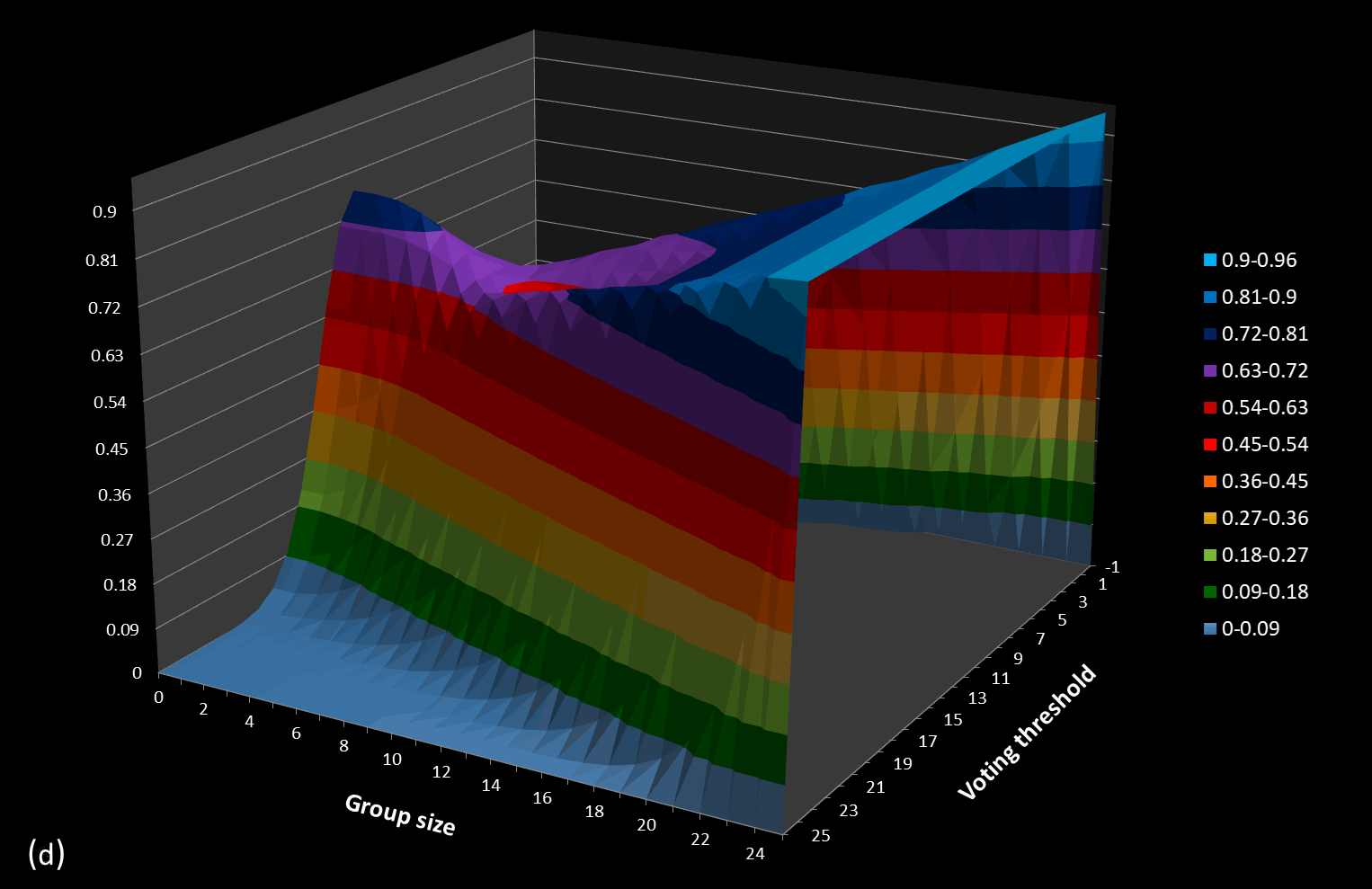}
}
\caption{\x{Expected one-step capital gain}The ECG of: (a) group members; (b)~individualists; (c)~difference between them;
(d)~\x{the }society's ECG. $I|G$ societies consist of $n=25$ agents under $\NOlt$ generator. 
\label{f:Group}}\end{center}\end{figure}

The performance of\x{ Group and Individualists} the two strategies under study\x{in single-group societies} is illustrated in Fig.~\ref{f:Group}.
\aB{These surfaces have been discussed in~\cite{CheMax21}.}

\section{Approved Transitions in Two-Component Societies}\label{s:Tra2Comp}

In this section, we study transformations of society aimed at increasing the welfare of\x{different categories of} agents. 
There are two types of changes of interest: structural and procedural.
Structural changes\x{transformations} of a single-group society are an increase or decrease in the group size. Procedural transformations are changes in the voting threshold. {\em Elementary transitions\/}\x{ transformations} are an increase or decrease in $g$ or $t$ by~$1.$\x{one.}

\subsection{Types of Transitions between Societies}\label{ss:Transi} 

\begin{figure}[th]\begin{center} 
\y{
\includegraphics[height=\wF cm]{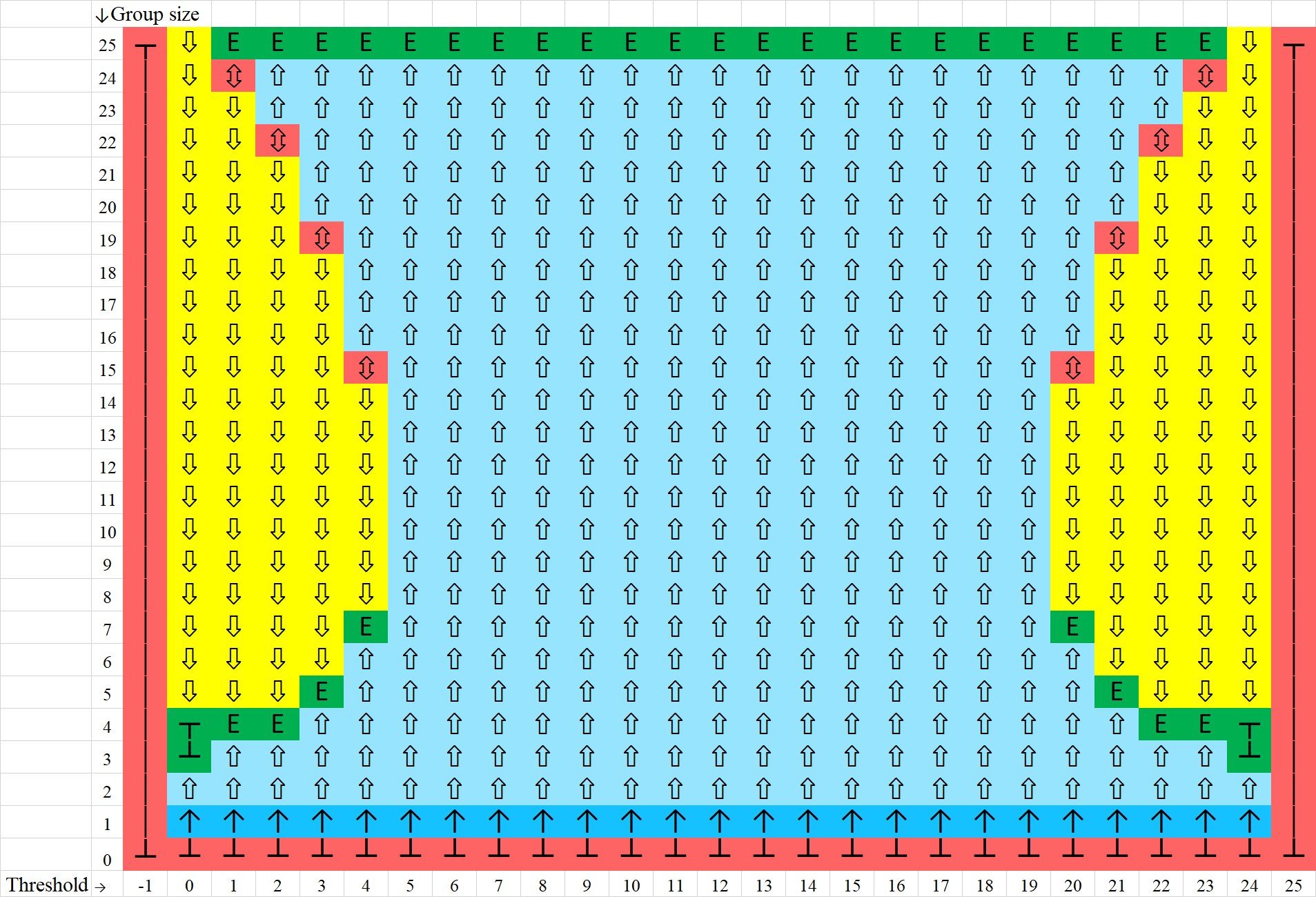}\\\vspace*{.8em}
\includegraphics[width=100mm]{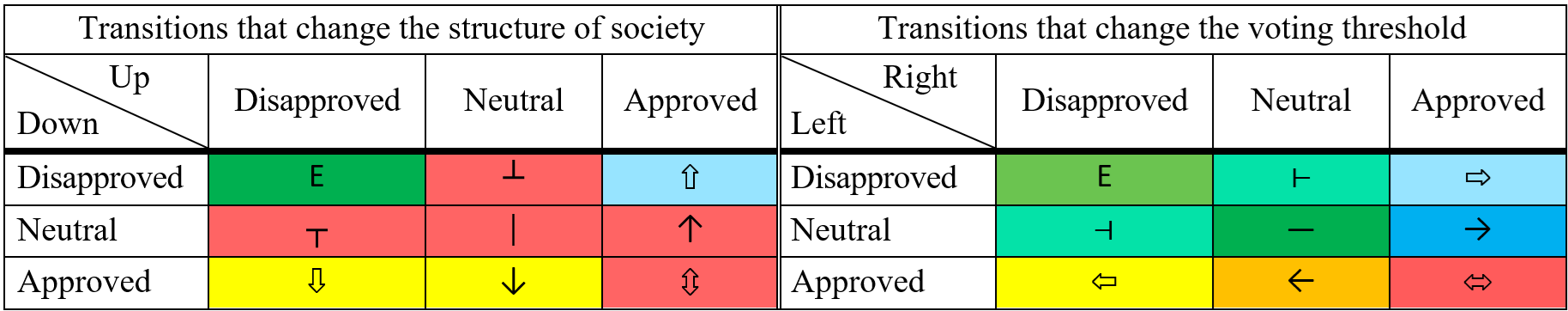} 
}
\caption{Changes in the degree of cooperation that benefit {\em defectors} (agents who change their strategy); symbols of transition between societies.
\label{f:T1}}\end{center}\end{figure}

We will analyze diagrams\x{ such as} similar to the one shown in Fig.~\ref{f:T1}. Each cell of a diagram corresponds to a society\x{ (denote it by $S$)} characterized by coordinates $t$ \aB{(voting threshold, on the $x$-axis)} and $g$ \aB{(group size, on the $y$-axis)}. 

The transition symbols\x{ used in such diagrams} are collected in Fig.~\ref{f:T1}. 
Each symbol in the left half\x{side} of the table encodes a {\em pair\/} of desirability/undesirability/neutrality indicators associated with transitions in both directions corresponding to a decrease or increase in the group size $g$ by one. 
The symbols in the right half of the table\x{ does the same} similarly refer to changes in the voting threshold~$t.$\x{ encodes desirability regarding} \aB{Desirability criteria vary for different types of agents and will be outlined below.}\x{ be\x{ determined} assessed according to different criteria.} A transition is ``neutral'' if the two societies being compared are equivalent according to the chosen criterion. 
The `-' symbol (not to be confused with double neutrality `$-$') means ``not applicable'' and is used when there are no agents of the corresponding category.

For clarity, the diagrams are colored. The colors correspond to symbols or their combinations.
\aB{They play a secondary role and are chosen informally.
The combination in which \textSFvi\ stands above \textSFvii\ is called the {\em equilibrium $2$-macrostate\/} and is marked in green like the equilibrium \textsf{E}.
}

\subsection{Cooperation vs Atomization}\label{ss:atomi}

\x{Let us find out }What changes in society are beneficial to\x{for} 1-agents, group members,\x{ or certain other categories, including} and agents changing\x{that change.4} their strategy? \x{Such}The latter ones\x{agents} will be called {\em defectors}.

The appearance of a defector changes the group size $g$ by one. \x{Recall that }This thus changes the arithmetic of the group members' strategy: they either begin to take into account the\x{counting a} defector \x{that}who joins the group, or stop\x{cease} taking into account\x{counting} the defector who leaves it.

A change in strategy is justified if it increases\x{ (or at least preserves)} the defector's ECG. 
Testing this with respect to society $S$ and society $S_{+1}$ (whose group has one member more than in $S$) amounts\x{reduces comes down.5} to checking this condition for both possible transitions between $S$ and $S_{+1}$. Namely, if the 1-agent's ECG in $S$ is smaller than the group member's ECG in $S_{+1}$, then this 1-agent benefits from joining the group (which transforms $S$ into $S_{+1}$) and an arrow pointing up to the $S_{+1}$ cell is placed into the $S$ cell. 
If, on the contrary, an 1-agent in $S$ has a greater ECG than the group member in $S_{+1}$, then the latter agent benefits from leaving the group, and an arrow pointing to $S$ is placed into the $S_{+1}$ cell. These comparisons\x{ indicated comparisons of agents' expected capital gains} of ECG values are based on Corollary~\ref{c:ECG}, for which the diagrams in Fig.~\ref{f:Group} may be useful. 

The $S$-cell contains one ``vertical'' symbol from the table in Fig.~\ref{f:T1} 
that aggregates the results of evaluating transitions from\x{comparisons of} $S$ to both $S_{+1}$ and $S_{-1}$ (whose group has one less member than the group in~$S$).

As already mentioned, the group size measures the level of cooperation in the society: $g=0$ corresponds to complete atomization; a society with $g=n$ is fully\x{completely1/4} cooperative.

The desirability diagram for defectors is shown in Fig.~\ref{f:T1}. Its arrows describe\x{show} the processes that can occur\x{in society} on the initiative of defectors, provided that both entry into the group and exit from it are permitted.
The main features of these processes are as follows.

If $t\in[5,19]$, then 1-agents in societies with group size of 2 to 24 benefit from joining the group and group members are not interested in leaving. This opens the way for\x{enables, makes it possible to implement} a ``snowball'' cooperation scenario \cite{CheLog10ARC,CheMax21}. 
When $g=1,$ an 1-agent benefits from joining a group, and nothing changes if the only group member becomes an 1-agent, since their group strategy is identical to the individualistic one. Every clique-like society ($g=25$)\x{ with $t\in[1,23]$} is an equilibrium:\x{ all agents are in the group,} leaving the group is unprofitable.

%
A general interpretation of\x{ these results} this diagram is as follow: In the most liberal ($t<5$) or conservative ($t>19$) societies, the members of a moderate-sized group may have an incentive to atomize; in societies closer to majoritarianism\x{ ones} (which means that decisions are made with a simple majority), 1-agents always have an incentive to choose cooperation. This is due to the fact that in liberal and conservative societies, the role of an individualist is highest: in the former, it is quite easy to implement her initiative, in the latter, a small faction can block changes that have high support. In some other cases we will also see that ``extremes meet'': in neutral environments, conservative and liberal societies have much in common that distinguishes them from majoritarian societies\x{, especially, in neutral environments}. 

Conservatism in decision-making is ``liberalism in reverse,'' that is, the freedom to block.
As a result,\x{In this regard, it is not surprising that} the dominant unvarnished\x{unembellished} news agenda of some \aB{real-world} conservative regimes is more about protests and reactions to them than about new ideas\x{initiatives}. 

\begin{figure}[th]\begin{center}
\y{
\includegraphics[height=\hFh cm]{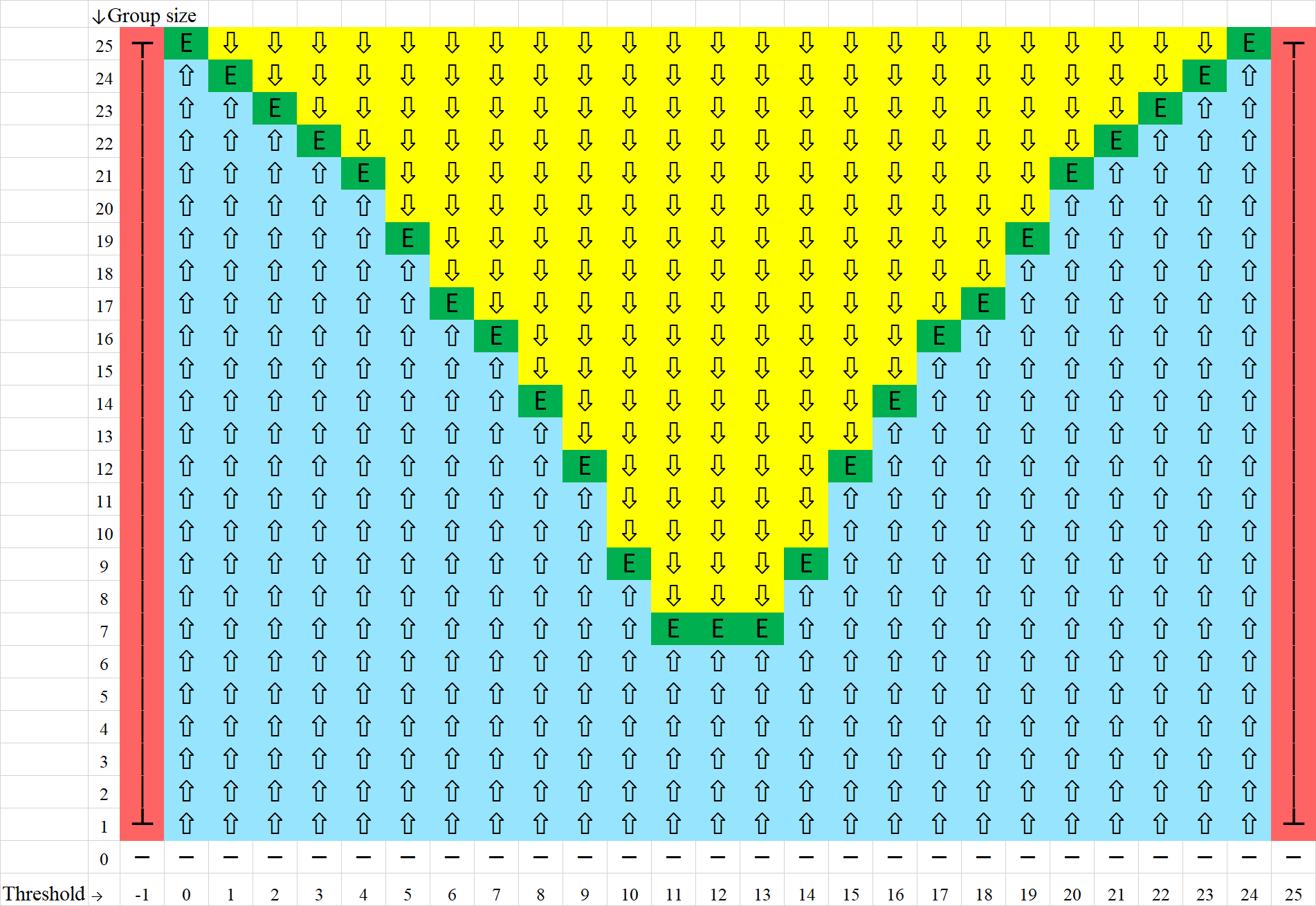}\;\includegraphics[height=\hFh cm]{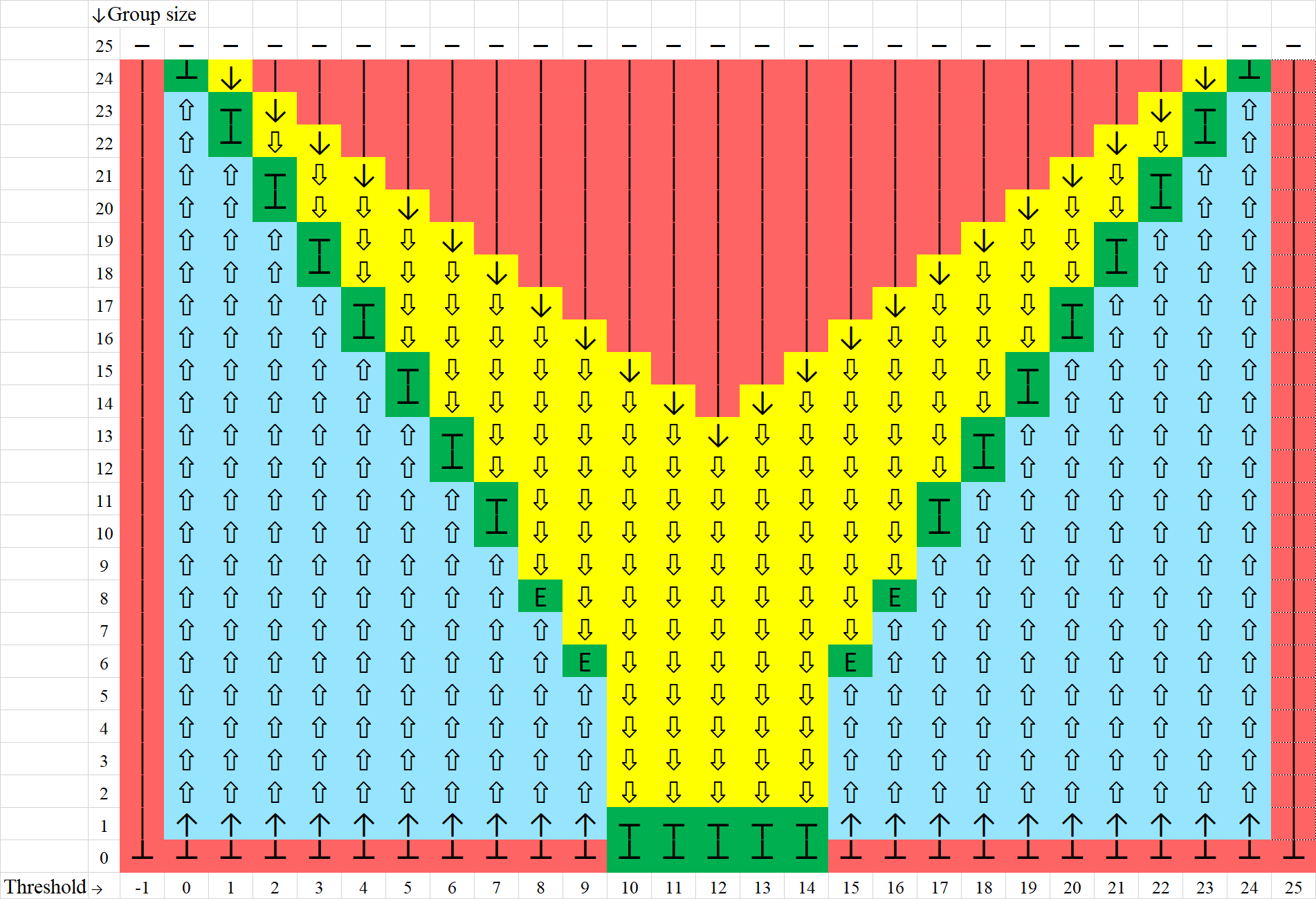}\\\ab
}
\caption{Changes in the group size\x{degree of cooperation} that benefit group members (a)\x{beneficial to} or 1-agents~(b).\label{f:T2}}\end{center}\end{figure}

Let us now find out which changes in the structure of society are beneficial to agents who {\em do not change\/} their strategies: group members (Fig.~\ref{f:T2}a) or 1-agents (Fig.~\ref{f:T2}b).

In the case of 1-agents, the diagram has the following main regions:
\begin{itemize}
\item triangular area of neutrality (\x{shown in }red), where all decisions are made by the group maximizing its capital gain, and the ECG of the 1-agents equals $\mu=0$; 
\item Y-shaped area (yellow), where 1-agents benefit from increasing their number\x{ and, accordingly, reducing the size of the group};
\item two blue areas where the absolute difference between the voting threshold and the simple majority threshold (we will call this difference the {\em specificity\/} of a voting rule or a society) is 3 to 12, and the size of the group does not exceed twice the specificity\x{ this absolute difference}. Here, 1-agents benefit from reducing their share in society.
\end{itemize}
Equilibrium states and $2$-macrostates\x{ comprising consisting of two societies} are marked in green. They\x{ locally} maximize\x{ Leaving them reduces} the ECG of 1-agents with respect to the group size for a fixed~$t$.

The desirability diagram for group members (Fig.~\ref{f:T2}a) \aB{has a simple description}: the closer the voting rule is to \x{a}the majority rule, the smaller the optimal (for group members) group size; for $11\le t\le13,$ \aB{this group size} is~$7$. At each $t,$ the group member's ECG is single-peaked.

\begin{figure}[th]\begin{center}
\y{
\includegraphics[width=\hEh cm]{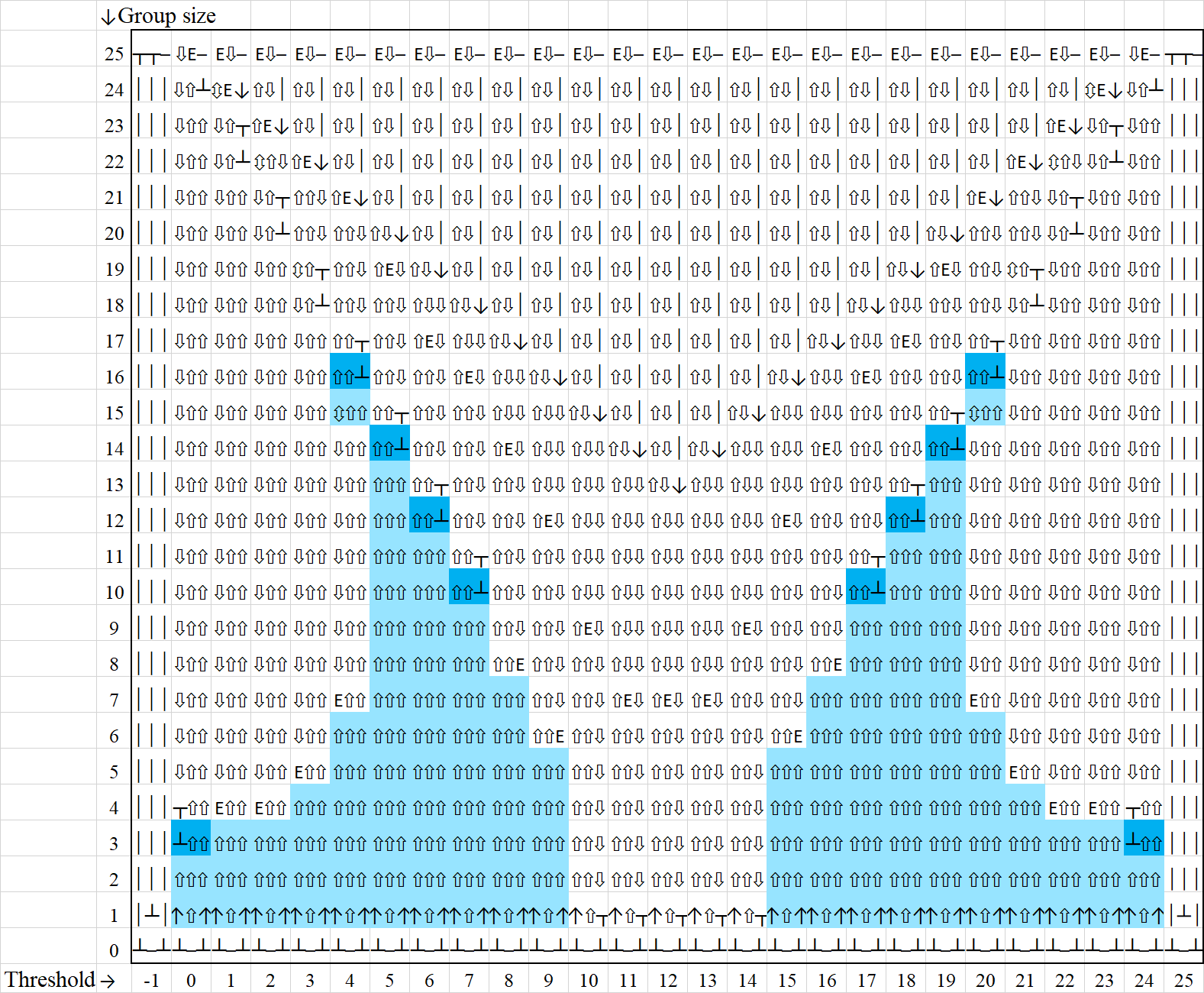}\includegraphics[width=\hEh cm]{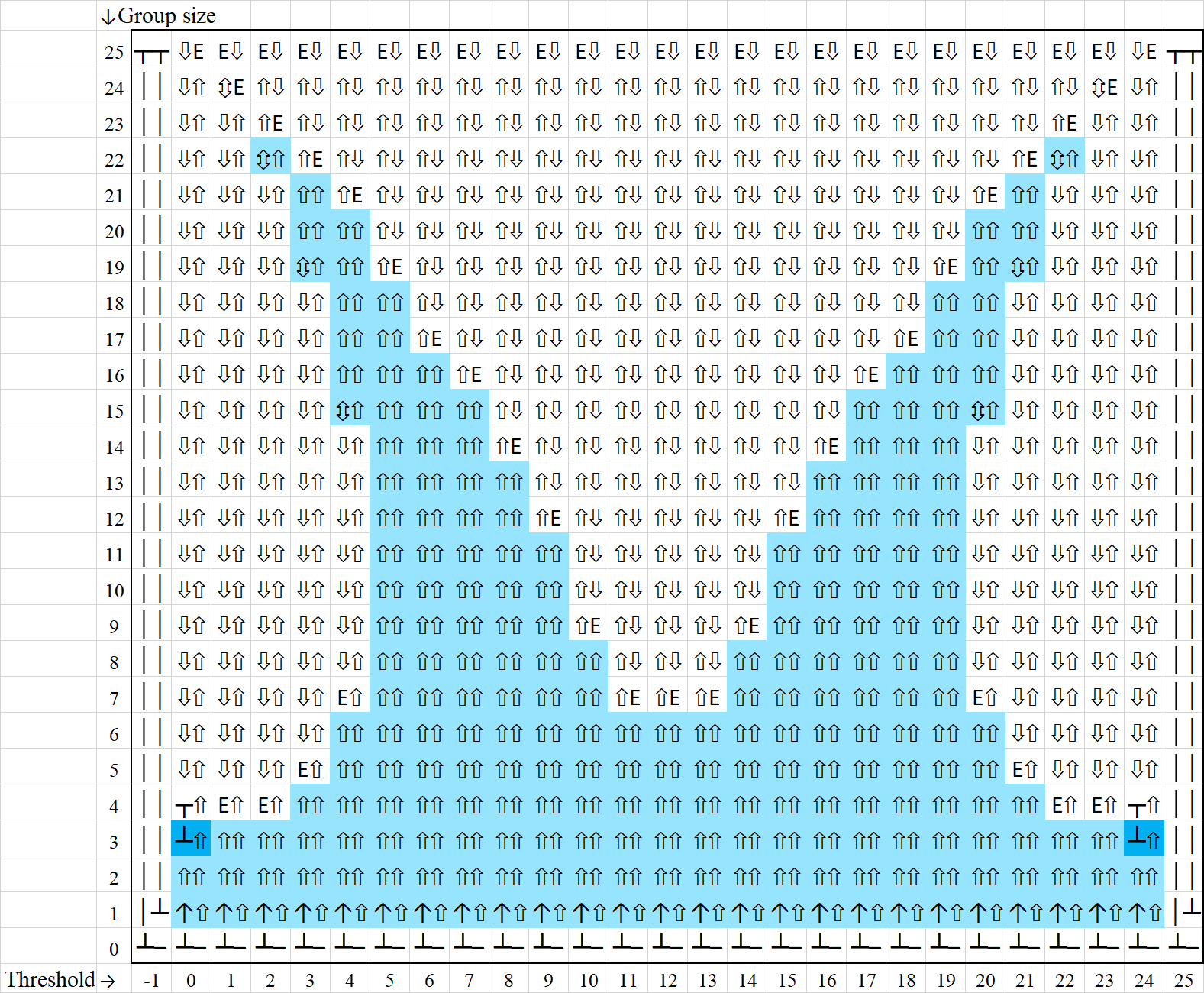}\\\ab\\ 
\includegraphics[width=\hEh cm]{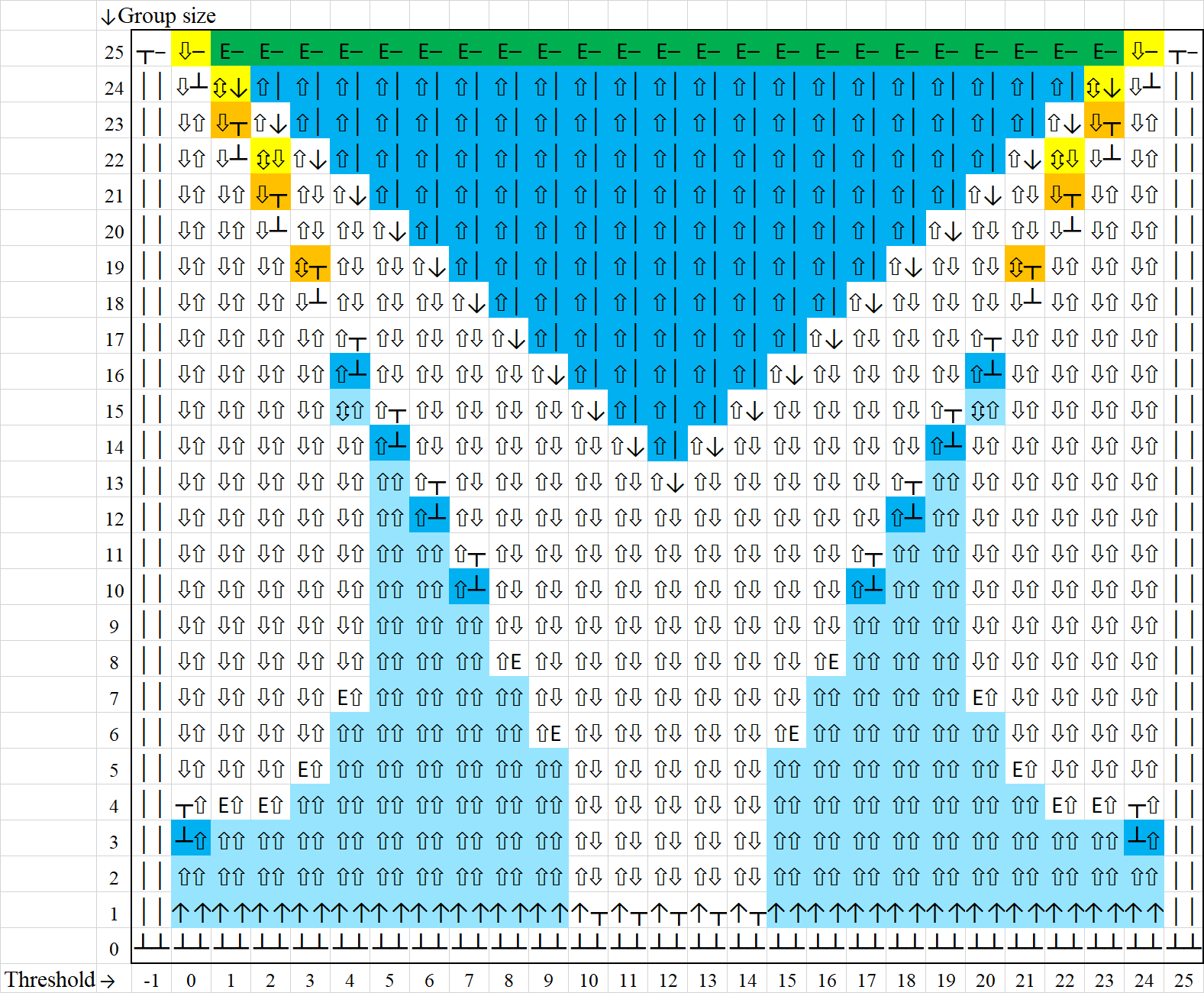}\includegraphics[width=\hEh cm]{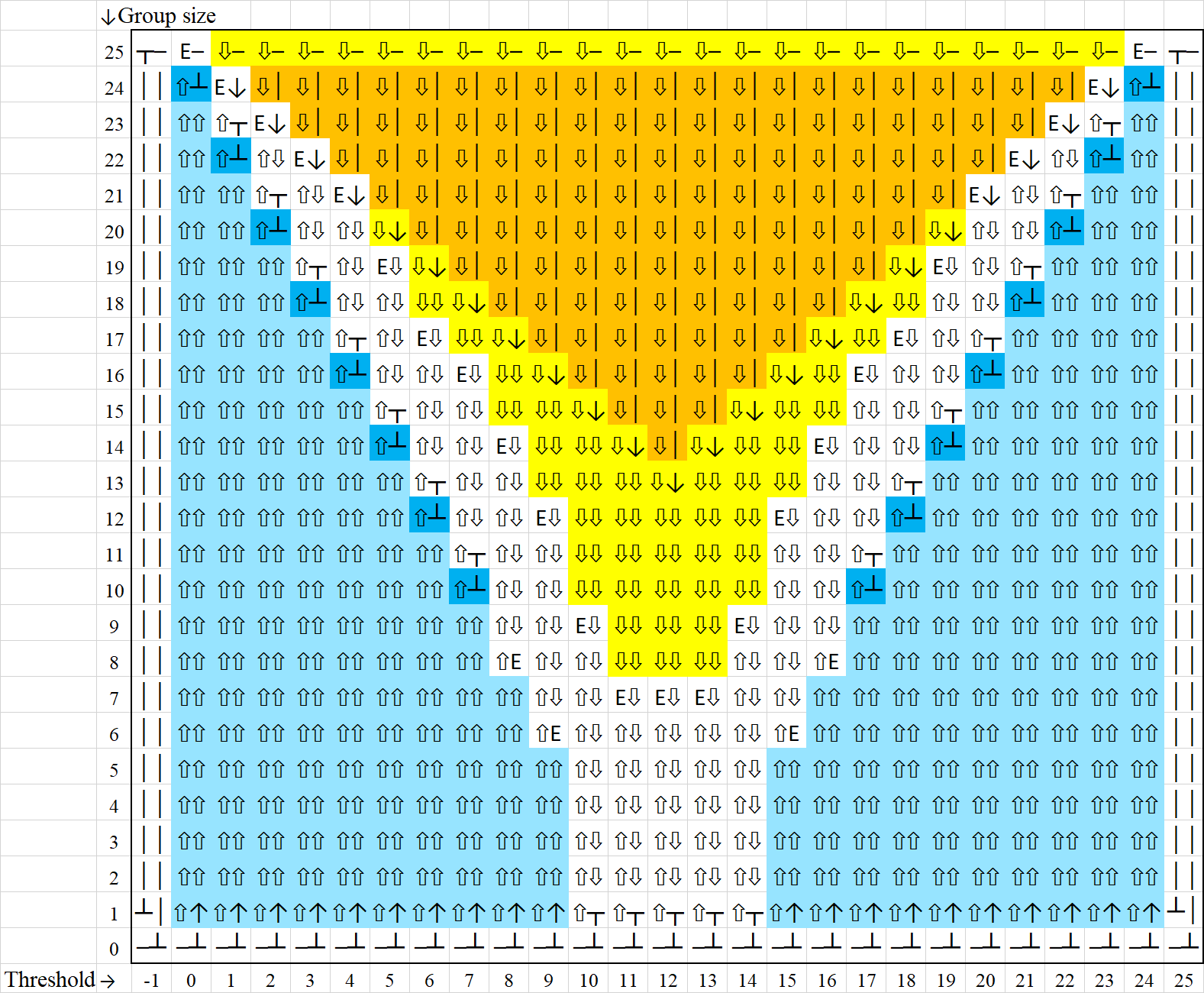}\\\cd
}
\caption{Approval of transitions by:
(a) a defector (the left symbol in each cell), group (the middle symbol), and 1-agents; 
(b) a defector (left symbol) and group;
(c) a defector (left symbol) and 1-agents; 
(d) the group (left symbol) and 1-agents.
\label{f:T4}\label{f:Concord}}\end{center}\end{figure}

Group members and 1-agents quite often benefit from the same changes in the degree of cooperation. In the diagram of Fig.~\ref{f:Concord}d, a decrease \aB{(resp., an increase)} in group size supported by these categories is marked in yellow (resp., in light blue); beige and dark blue colors respectively indicate a decrease and an increase with neutrality of one category and support of the other.

The areas of agreement of one\x{ each either} of these categories and the defectors\x{ (Fig.~\ref{f:T1})} are shown in Figures~\ref{f:Concord}b and~\ref{f:Concord}c; Fig.~\ref{f:Concord}a presents the transitions approved by all agents. 

All transitions\x{ approved by all agents are shown in Fig.~\ref{f:Concord}a.} of common agreement expand the group and thus fit into the ``snowball  of cooperation'' scenario. This scenario has practical applications, among which we note\x{mention} \cite{Kovalev10a}, where it is described as follows: ``The initial union [of actors] should be ready to accept new members, and those should wish to join this association. The reason for this desire can only be the obvious and significant advantages of the new status, and practical ones at that.'' In Fig.~\ref{f:Concord}a\x{our case}, transitions approved by everyone are implemented at voting thresholds that differ from the\x{ simple} majority threshold by at least 3, while the group size belongs to \aB{the interval} $[3,16].$\x{ limited by numbers from 3 to~16.} Transitions for which one of the categories is neutral are marked in dark blue. Together with the consensus transitions, they constitute\x{make up5/7} 22\% of non-trivial societies.

\subsection{Democratic and Mafia Groups}

So far we have assumed that each 1-agent can\x{freely} enter the group, and each group member can leave it by becoming an 1-agent. Such a group can be called {\em open}. 
However, many real-world groups, like\x{such as} political parties, behave differently. They do not accept everyone, but only those whose entrance\x{joining they consider} seems beneficial for the group\x{mselves}. Any member has the right to leave the group. In \cite{Kovalev10a},
a group of this type is described as follows: ``Joining the association... should be conditioned by a meticulous assessment of the candidates' readiness, and exit should be unhindered.''

\begin{figure}[ht]\begin{center}
\y{
\includegraphics[width=\wFH cm]{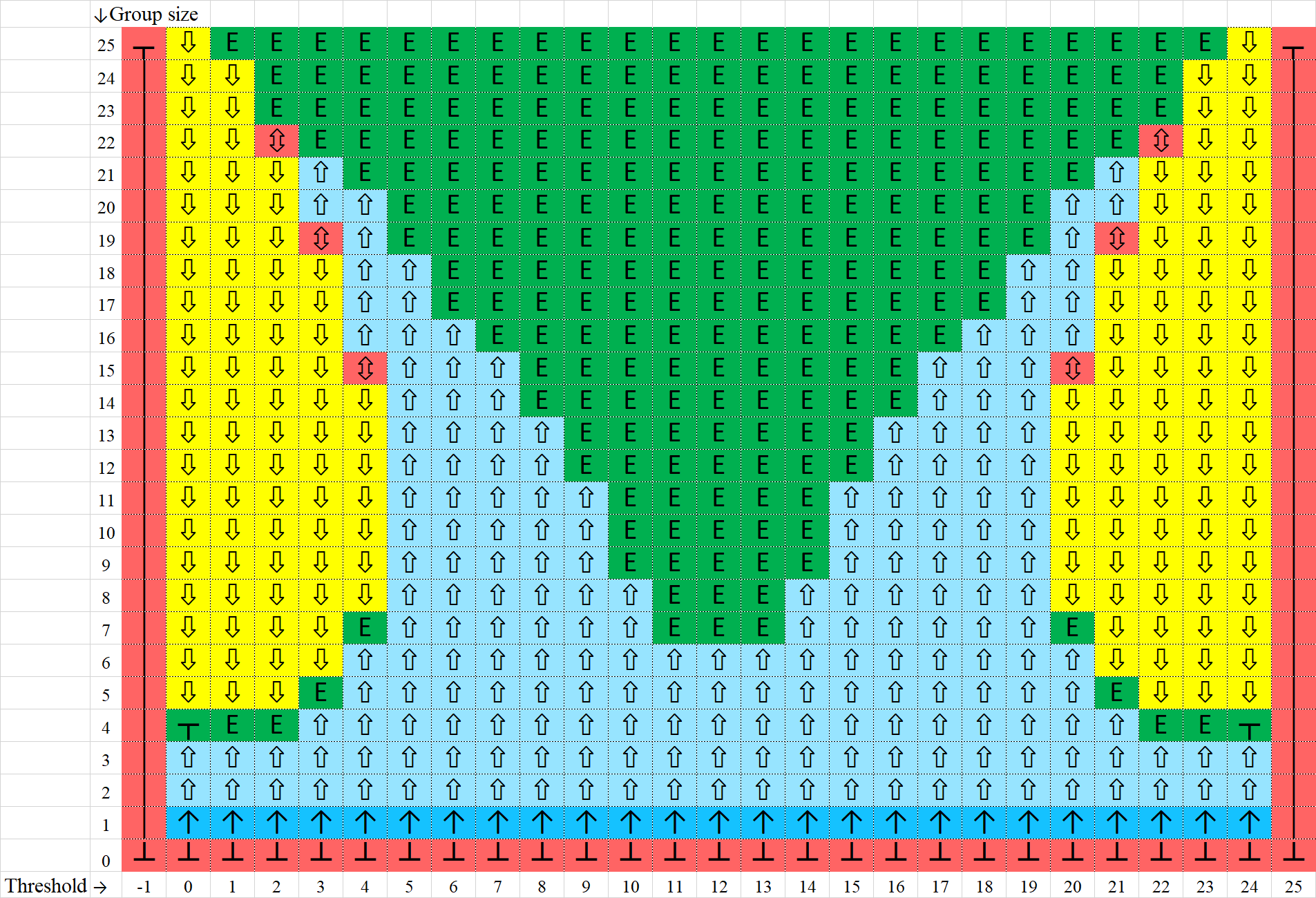}\;\includegraphics[width=\wFH cm]{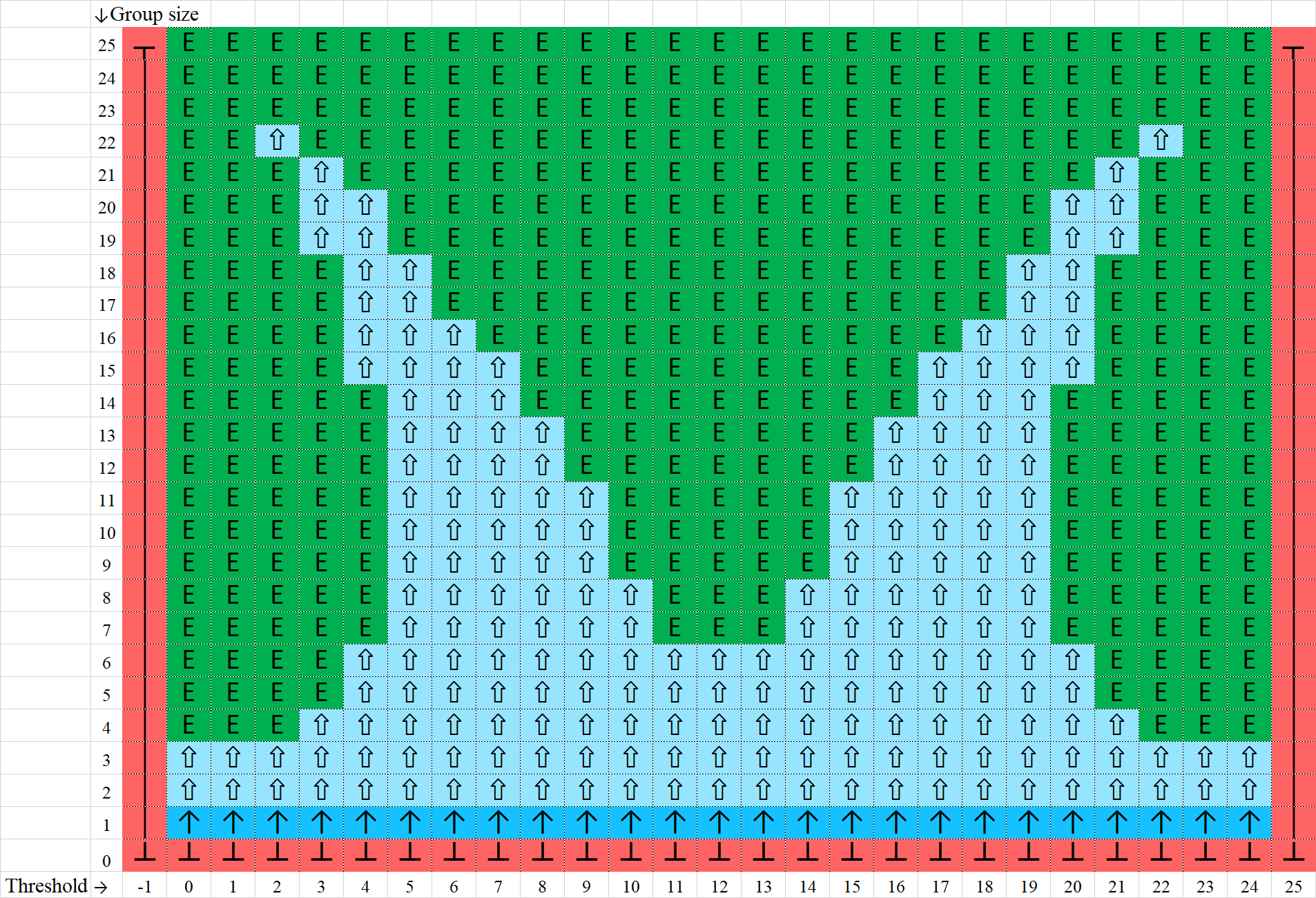}\\\ab
}
\caption{Approved changes in the degree of cooperation in \aB{the presence} of a democratic group~(a) or a mafia group~(b).\label{f:T5}\label{f:StruDemoc}}\end{center}\end{figure}

Fig.~\ref{f:StruDemoc}a shows the transition diagram in the case where a necessary condition for joining the group, in addition to the benefit of its new member, is the consent\x{agreement} of the group, which is given if the group expansion increases the ECG of the current group members. The justification\x{rationale} for this \aB{conditioning}\x{sanction} in the ViSE model may be that when accepting a new member, the group must expand\x{extend,widen<} the concept of ``group interest'' by taking into account the interests of the defector\x{applicant}, which reduces the degree of consideration of the interests of the current\x{present,existing} members.\x{previous participants is reduced.}

Authorized entry and voluntary exit are consistent with the practice of democratic organizations, as opposed\x{in contrast} to mafia-type associations\x{organizations}. Therefore, we will call this type of group {\em democratic}.
If the ECG of the entering 1-agent remains the same\x{does not change}, then her entry to the group will be considered approved if it increases the group members' ECG, prohibited if decreases, and neutral if preserves.\x{it remains unchanged.} This means the goodwill of the 1-agent: in the case of her own neutrality, she favors an action that benefits her possible partners. However,\x{At the same time, the criterion of} their benefit is lexicographically subordinate to the defector's own benefit.\x{ criterion of the benefit of the agent.}

With a democratic group, the more ``specific'' (\aB{conservative or liberal}\x{further from the majoritarian}) the society\x{voting rule}, the higher the minimum level of cooperation that ensures its\x{the} structural stability (equilibrium)\x{ of society}.

A diagram for a group that blocks {\em all\/} transitions that are unfavorable for its members (we call such a group a {\em mafia}\x{, or a {\em mafia group}}) is presented\x{shown} in Fig.~\ref{f:StruDemoc}b.

The conclusion is that although a mafia group may be interested in reducing its size (Fig.~\ref{f:T2}a), it never approves of a member's initiative to leave it. Like a democratic group, it has the greatest prospects for expansion in moderately liberal or\x{and} moderately conservative societies.

\subsection{Liberalism vs Conservatism}\label{ss:LibCons}

\begin{figure}[ht]\begin{center}
\y{
\includegraphics[width=\hEh cm]{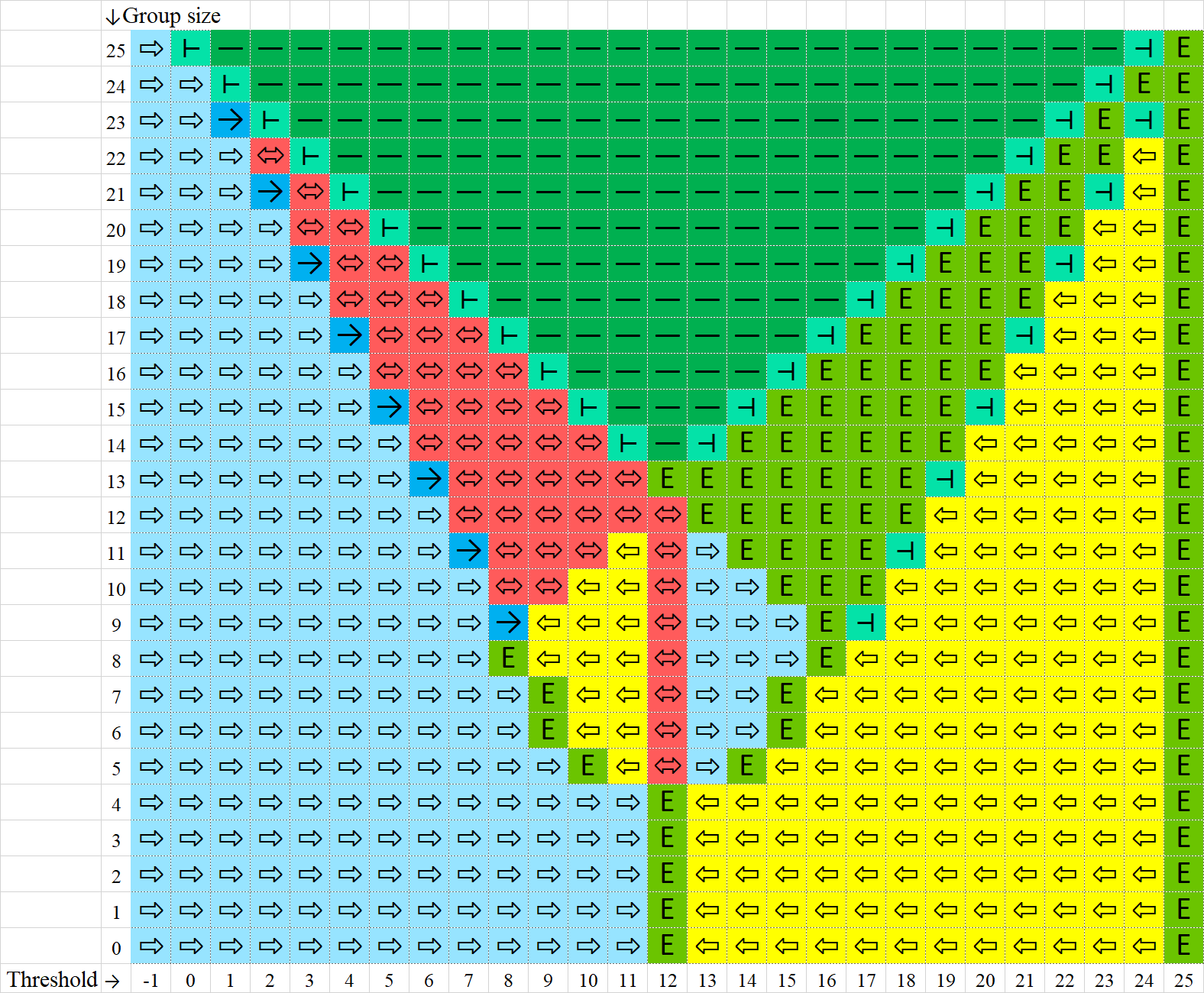} \includegraphics[width=\hEh cm]{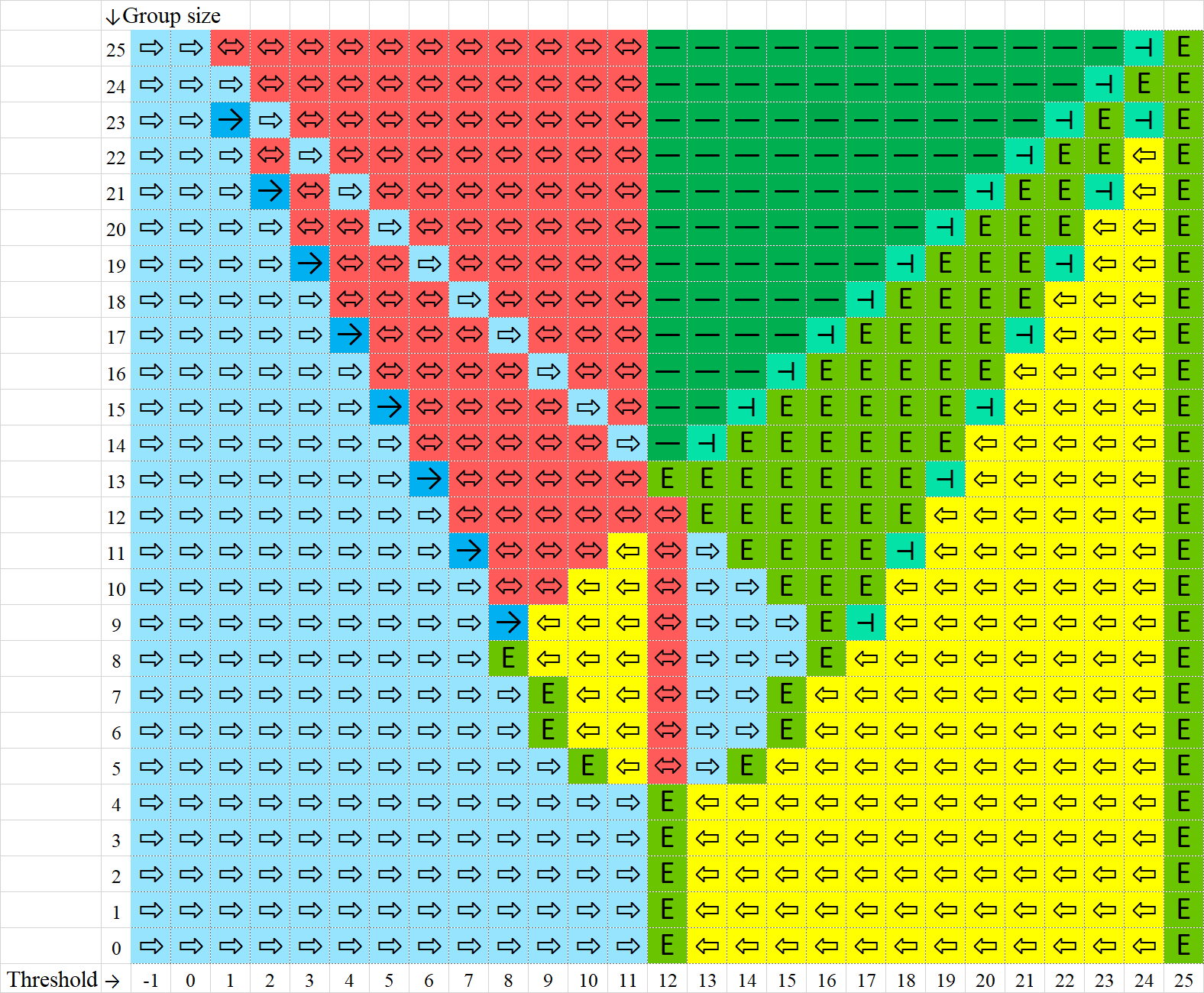}\\\ab    
}
\caption{Approved changes in the voting threshold made by voting with the current threshold. Abstention votes: (a) do not change\x{increase} support; (b) are counted in half\x{are taken into account with a coefficient of $0.5$; are counted by a factor of 0.5}.\label{f:T7}\label{f:ChTh}}\end{center}\end{figure}

\x{It is natural to}We assume that the decision to change the voting threshold by one is made by voting with the current threshold. The change is supported by agents to which it brings an increase in the ECG.
Diagrams of transitions consisting in changing the voting threshold are presented in Fig.~\ref{f:ChTh}.

If \aB{a proposal preserves} the agent's ECG, then this agent abstains; the votes of such agents do not contribute to the number of votes ``for'' (Fig.~\ref{f:ChTh}a) or are taken into account with a coefficient of $0.5$ (Fig.~\ref{f:ChTh}b). \aB{In the latter case,} an abstention provides $0.5$ votes ``for'' and $0.5$ votes ``against''. \x{These rules}The corresponding diagrams have a minor difference\x{ results of applying these methods of accounting for abstainers differ slightly}: it concerns only the approval of certain transitions between the states equivalent in Fig.~\ref{f:ChTh}a. Let us consider the former\x{ first} rule.

\x{An interpretation of these results is}
\x{We conclude}\aB{The diagram in Fig.~\ref{f:ChTh}a shows} that radically liberal or conservative societies \aB{tend to become less ``specific''}.\x{ evolve\x{develop} in the direction of majoritarian ones: the more so, the lower the degree of cooperation.} However, they become majoritarian only if the group size is less than~$5.$ If the degree of cooperation is high, this movement soon ceases.

Interestingly, majoritarian and near-majoritarian societies with a moderate degree of cooperation ($g\in[5,11]$) are characterized by a centrifugal tendency: they tend to increase their specificity. \x{ become more ``specific,'' move away from majoritarian ones. i.e., more conservative or liberal.} 
This is because of the individualists: a small proportion of them, together with the group, can accept an proposal that is advantageous to them in a relatively liberal society or reject an unfavorable proposal in a relatively conservative one; in a majoritarian society these possibilities are absent.

With greater cooperation, \aB{liberal societies fall into the red area\x{region} of bidirectional transitions (``dynamic equilibria''), which then gives way to the green region of equilibrium macrostates, while conservative societies stop on the border} of the light green area of equilibria.

\subsection{Combinations of Structural and Procedural Transitions}\label{ss:BothTrans}

\begin{figure}[ht]\begin{center}
\y{
\includegraphics[width=\wFH cm]{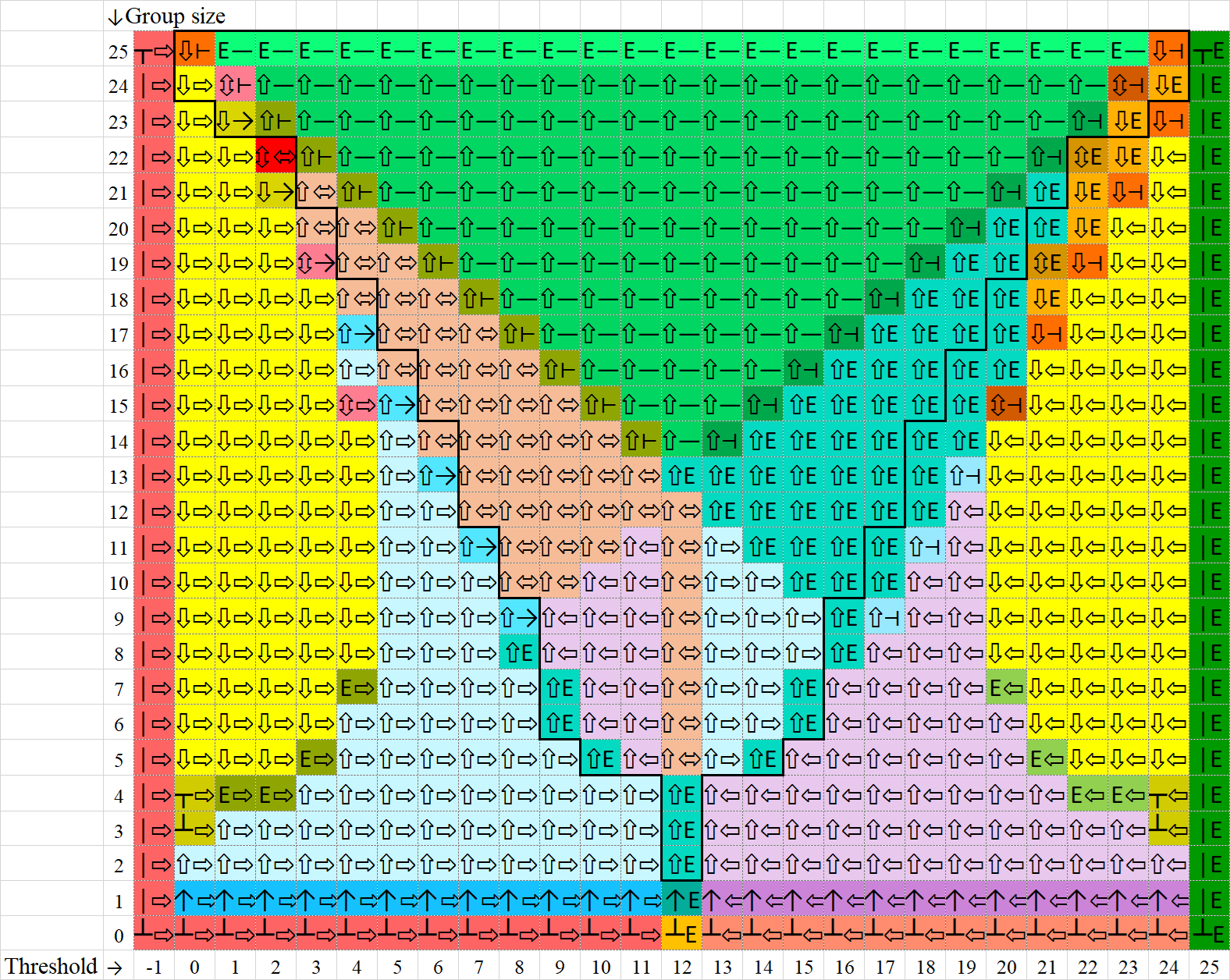}\;\includegraphics[width=\wFH cm]{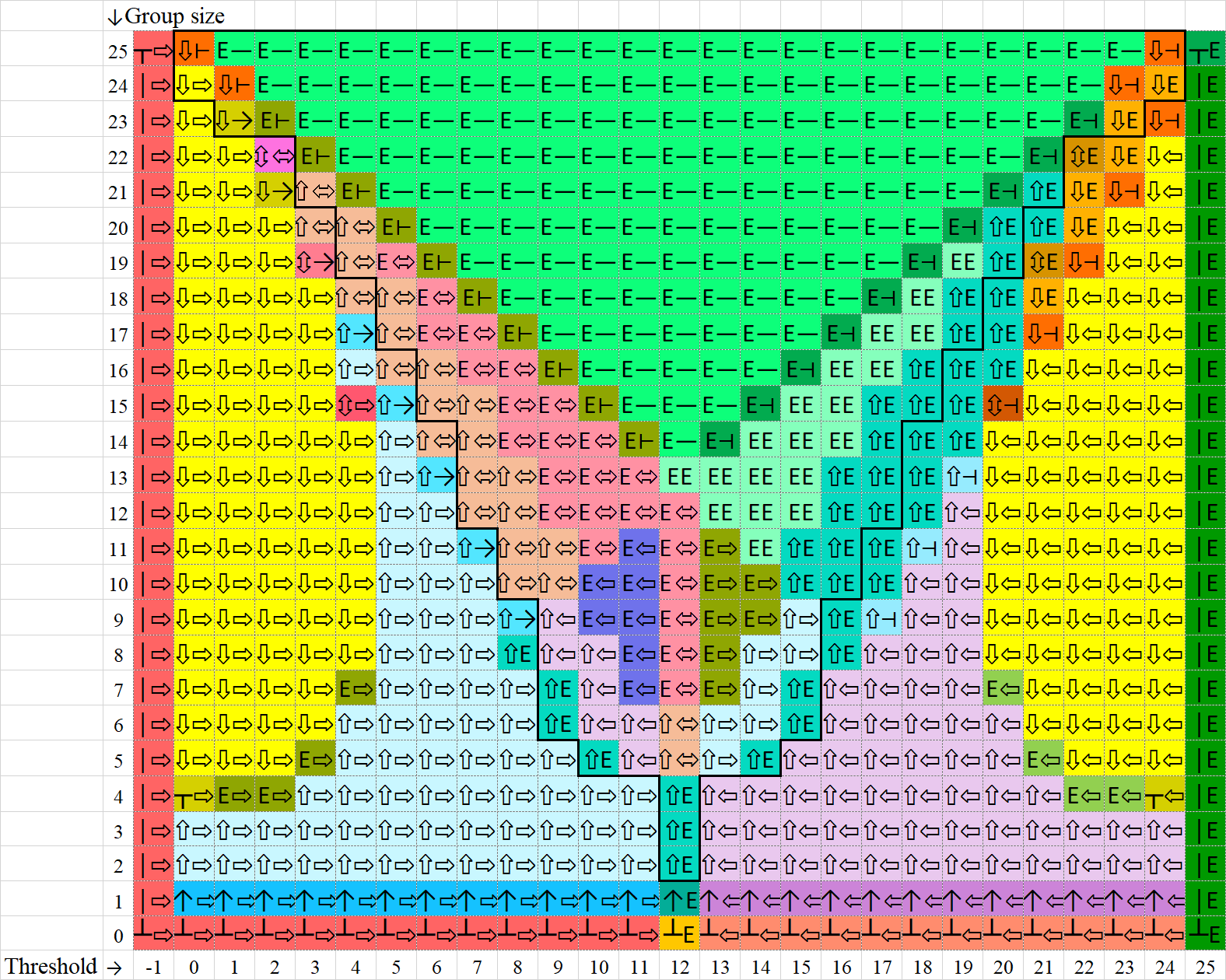}\\\ab    
}
\caption{\aB{Approved} structural and procedural transitions in the case of: (a)~an open group; (b)~a democratic group.\label{f:T8}\label{f:2ChO}}\end{center}\end{figure}

The most interesting dynamics are realized when both structural and procedural transitions are allowed.
Then, to analyze the possible evolution, \x{you}we need to combine the diagrams shown in Figures~\ref{f:T1}, \ref{f:StruDemoc}a,b, and~\ref{f:ChTh}a. The results are shown in Fig.~\ref{f:2ChO}a (the case of an open group) and Fig.~\ref{f:2ChO}b (for a democratic group). The diagram for a mafia group is presented below in Fig.~\ref{f:CycCapi}b.
\aB{Since many of the $81$ combinations of the transition symbols (Fig.~\ref{f:T1}) appear on the combined diagrams, the colors (which play a secondary role) only partially correspond to those attached to the symbols.}
The black stepped line shows the boundary of the Pareto set.\x{ (Fig.~\ref{f:Pareto}), which narrows as the group size decreases.}

In the first case (Fig.~\ref{f:2ChO}a), there are two equilibrium macrostates: $t=25$ (rejection of all proposals for any group size) and $g=25$ for $1\le t\le23$ (all agents are in the group).

The main difference \aB{between the democratic and open group cases} 
is the blocking of group's expansion\x{the prohibition of increasing the group} 
(upward movement) in the large green triangle in Fig.~\ref{f:StruDemoc}a when the group is democratic. This leads to the appearance of
(a)~equilibrium macrostates at $14\le g\le24$ and
(b)~a region of equilibrium states, the corner points of which are $(12, 13)$, $(14, 11),$ and $(19, 19)$.
For society as a whole, the case of a democratic group is often less beneficial: by prohibiting the unprofitable entry of 1-agents, the group limits the growth of society's capital.

\section{Combined Evolution Scenarios}\label{s:Combi}

By {\em combined scenarios\/} ({\em routes\/}) we mean the scenarios that include changes in both the structure of society and the voting \aB{threshold}, where all changes are approved using the relevant rules discussed above.
Can such scenarios be cyclic\x{al1/4}? 

\begin{figure}[ht]\begin{center}
\y{
\includegraphics[height=\hEEh cm]{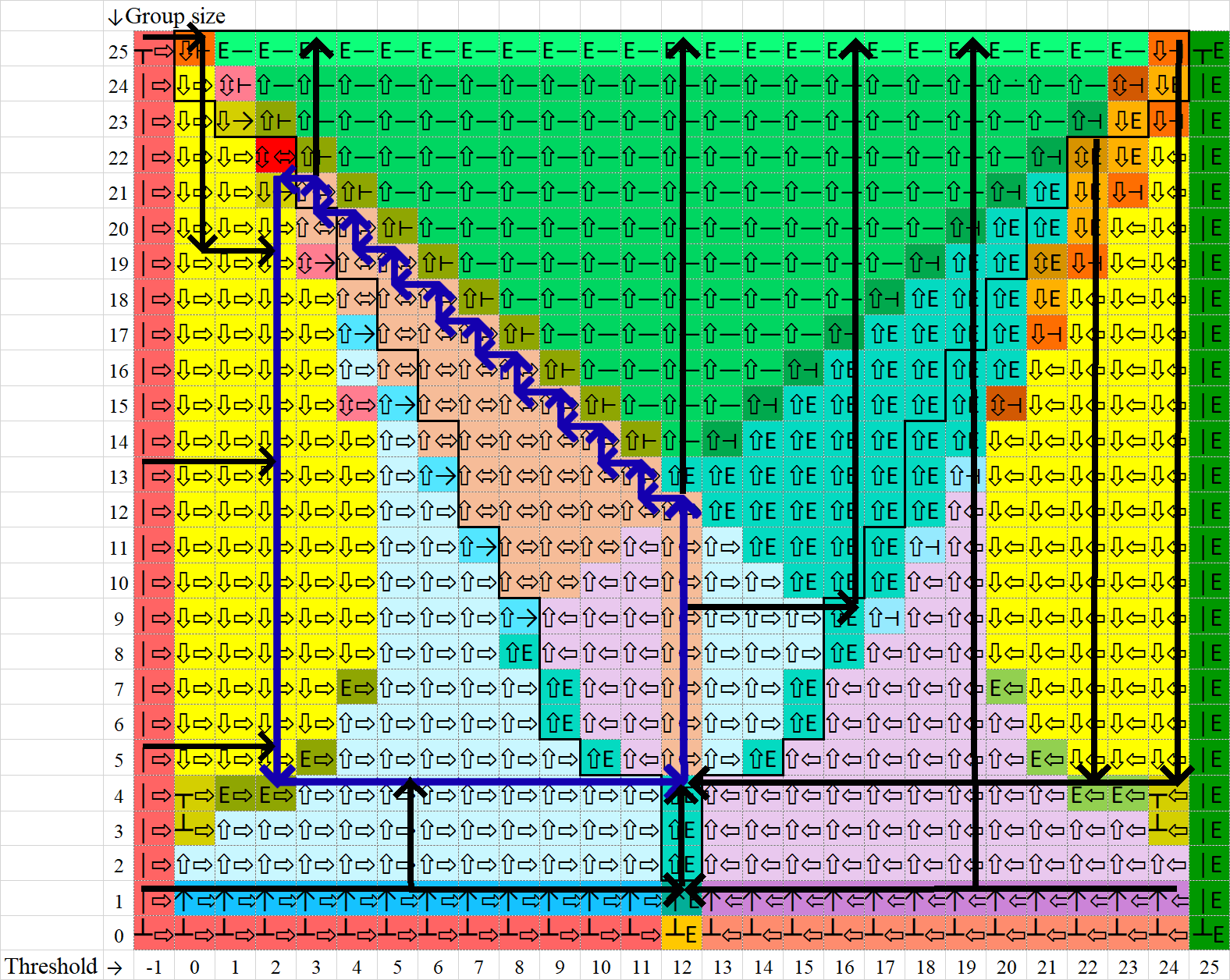}\includegraphics[height=\hEEh cm]{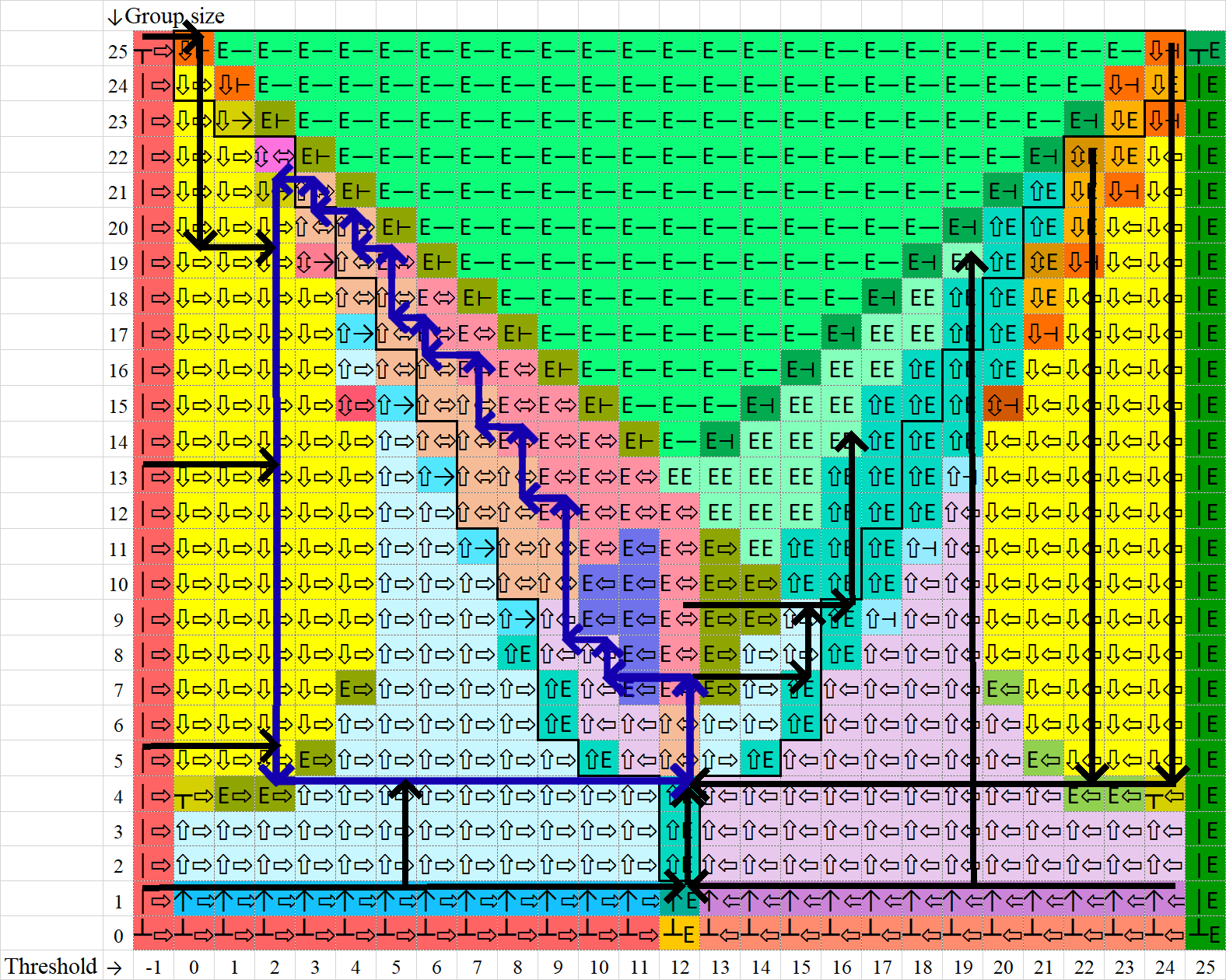}\\\ab
}
\caption{Combined scenarios with: (a)~an open group; (b)~a democratic group.\label{f:T12}\label{f:Combi}}\end{center}\end{figure}

The answer is yes;\x{ one example,} \aB{a dark blue counterclockwise cycle} is shown in Fig.~\ref{f:Combi}a. 
\aB{All states in the region bounded by this cycle plus the $(2,22)$ state are mutually accessible; the corresponding societies are majoritarian or liberal.
This trapezoidal region is described as $t\in[2,12],$ $g\in[4,22-t].$ It is accessible from different\x{ various} parts of the `universe'\x{diagram}. Exceptions are some conservative societies and the large green upper triangle.

\x{On the other hand,}Note, however, that conservative societies with low cooperation ($g\in[0,4]$; recall that $g=0$ and $g=1$ are equivalent) or extreme conservatism ($t\in[22,24]$) without extremely large ($g>22$ for $t=22$ or $g=25$ for $t=23$) groups have paths to that trapezoid. Typical evolution paths are shown by black arrows.
}

\begin{figure}[ht]\begin{center}
\y{
\includegraphics[height=\hFh cm]{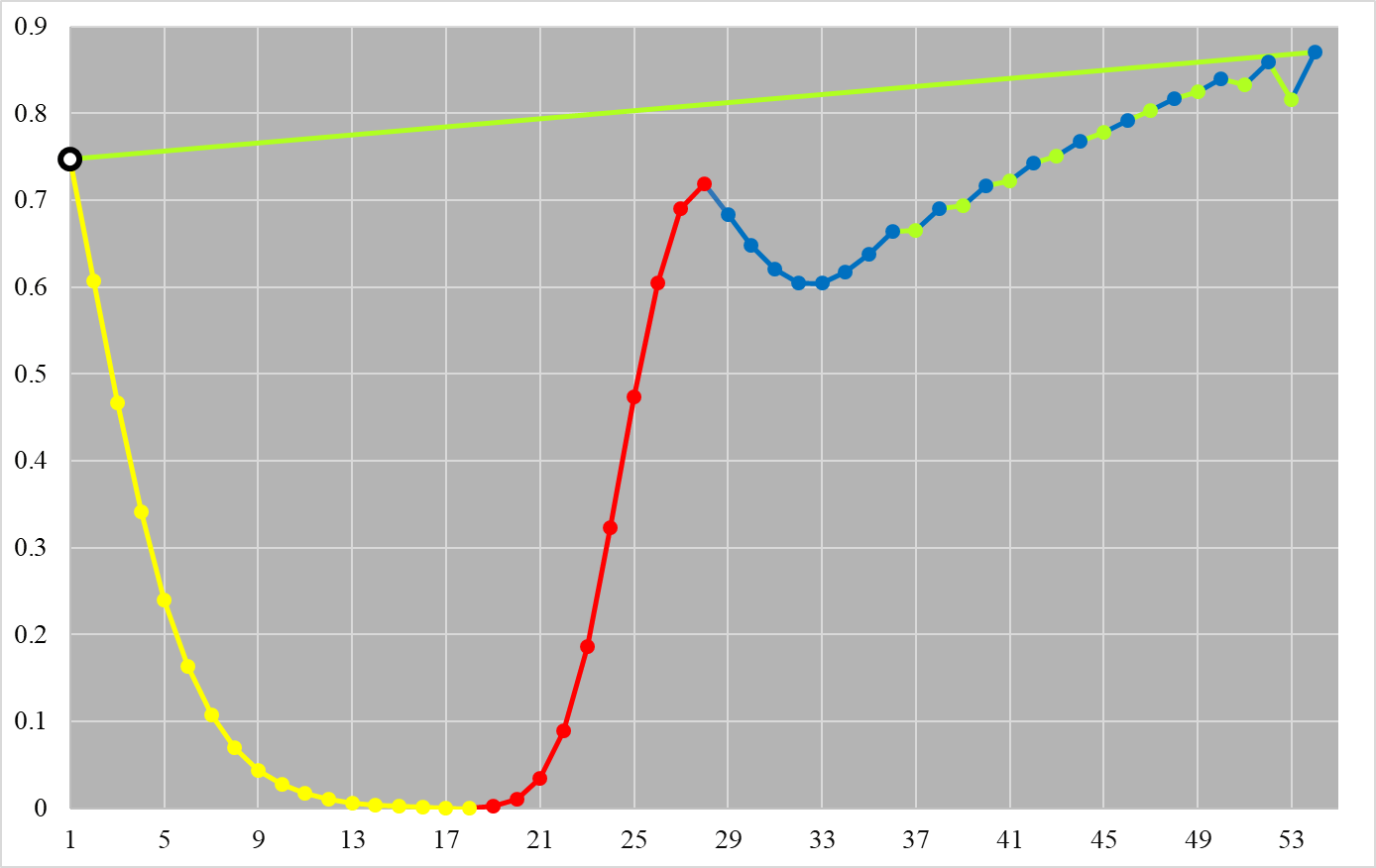}\includegraphics[height=\hFh cm]{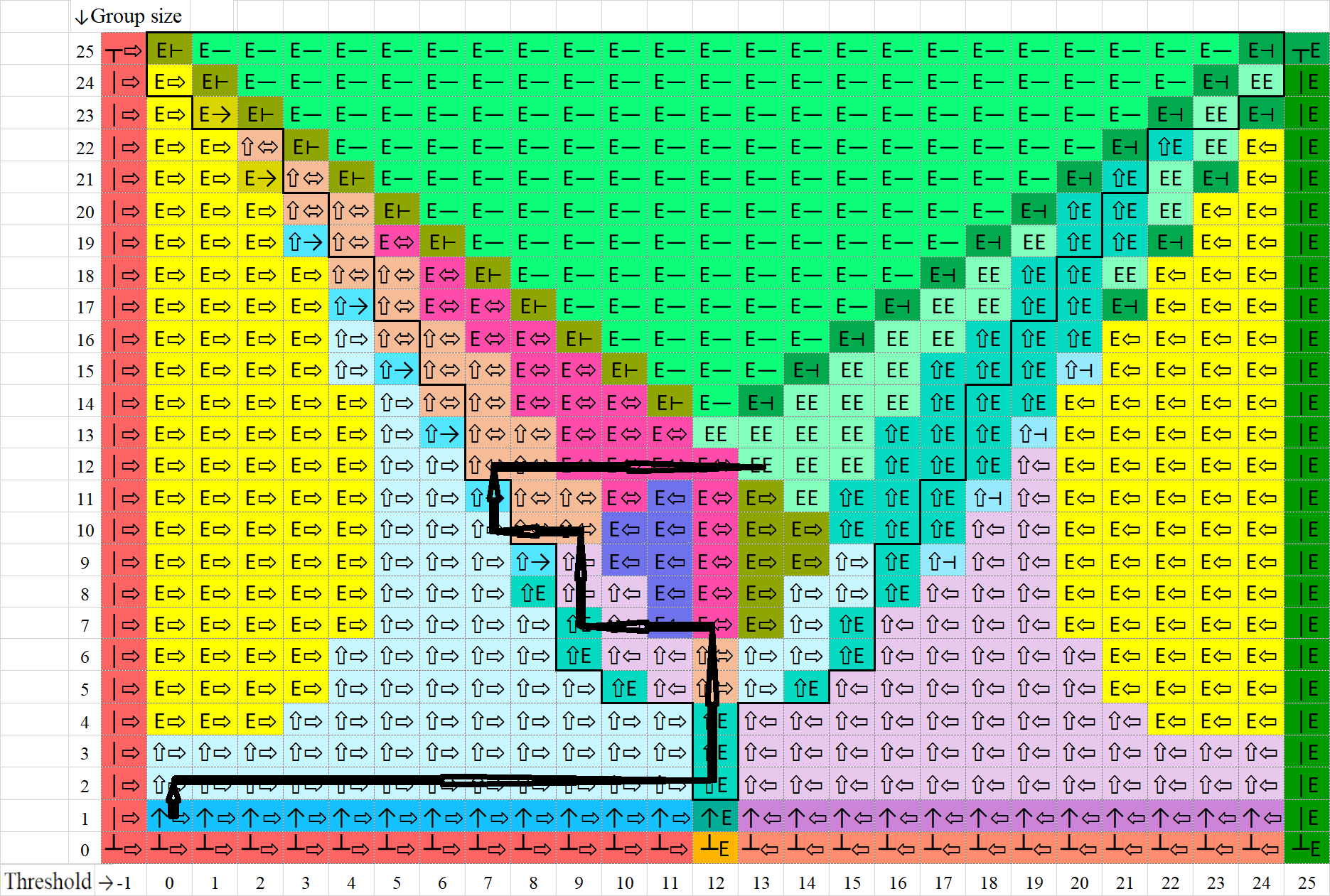}\\\ab 
}
\caption{(a)~ECG of the societies on the cyclic route shown in Fig.~\ref{f:Combi}a 
depending on\x{against} the step number starting from $(t=2,\,g=21)$ 
at $\sigma=12$;
growth of: atomization (yellow); cooperation (blue); liberalism (green); approaching majoritarianism (red)\x{the $x$-axis is the society number}.
(b)~A combined scenario for societies with a mafia group.
\label{f:T13}\label{f:CycCapi}}\end{center}\end{figure}

How does the ECG of society change along the considered cyclic route? Its dependence\x{ of the ECG for a society} on the \aB{step} number starting from $(t=2,\,g=21)$ is shown in Fig.~\ref{f:CycCapi}a.
\aB{Along this cycle, any increase in atomization (yellow) reduces the ECG, any move toward majoritarianism (red) increases it, while the impact of increased cooperation (blue) or increased liberalism (green) is ambiguous.} 

\aB{In the case of a democratic group (Fig.~\ref{f:Combi}b), some `upward' transitions are prohibited\x{forbidden-more personal} by the group. A counterclockwise cycle still exists, but the region of mutually accessible states is smaller.
}

Mafia groups do not allow their reduction at the initiative of the participants, so no cyclic scenarios are possible.
However, there are scenarios with a non-monotonic change in the threshold, one of which is \aB{shown} in Fig.~\ref{f:CycCapi}b. 

\section{Conclusion}

Within the framework of the ViSE model, we studied the evolution of a two-component society, which consists of a change in the level of cooperation (group size~$g$) and the degree of conservatism (voting threshold~$t$).
\aB{We considered societies with $n=25$ members; the main qualitative findings hold for increasing or moderately decreasing~$n$.}

Non-trivial evolution patterns \aB{were identified. In the case of an open group, all societies 
belonging to the trapezoid $t\in[2,12],\,g\in[4,22-t]$ participate\x{is involved} in cyclical patterns with all other societies of this region.
They are accessible from conservative societies with low cooperation or very high conservatism without extremely large groups and from other liberal or majoritarian societies not belonging to the upper green triangle in Fig.~\ref{f:Combi}a.
For societies with democratic groups, cyclical patterns exist, but some transitions that expand the group are prohibited.
A mafia group never lets its members go when they want to, but there are patterns with a non-monotonic change in the threshold~$t$.

The insights derived from parsimonious models, such as the ViSE framework, facilitate the elucidation of conditions that potentially govern the \x{manifestation}occurrence of similar phenomena in empirical contexts.}

\section*{Acknowledgment}

This work was supported by the Israel Science Foundation (grant No. 1225/20) and by the European Union (ERC, GENERALIZATION, 101039692).

\bibliographystyle{splncs04}       
\bibliography{all2} 

\begin{thebibliography}{10}
\providecommand{\url}[1]{\texttt{#1}}
\providecommand{\urlprefix}{URL }
\providecommand{\doi}[1]{https://doi.org/#1}

\bibitem{Afonkin21tax}
Afonkin, V.A.: Tax incentives for prosocial voting in a stochastic environment.
  Control Sciences  \textbf{1}(1),  53--59 (2021). \doi{10.25728/cs.2021.1.6}

\bibitem{Aizerman81e}
Aizerman, M.A.: Dynamic aspects of voting theory (survey). Automation and
  Remote Control  \textbf{42}(12),  1664--1675 (1981)

\bibitem{AfoChe25}
Chebotarev, P., Afonkin, V.: Majority voting is not good for heaven or hell,
  with mirrored performance. Annals of Operations Research  (2025), to appear

\bibitem{CheMax21}
Chebotarev, P., Maksimov, V.: Two-component societies in the {ViSE} model: How
  the level of cooperation and voting threshold affect the capital dynamics.
  Large-Scale Systems Control  \textbf{93},  51--88 (2021).
  \doi{10.25728/ubs.2021.93.2}, ({in Russian})

\bibitem{Che06ARC}
Chebotarev, P.Y.: Analytical expression of the expected values of capital at
  voting in the stochastic environment. Automation and Remote Control
  \textbf{67}(3),  480--492 (2006). \doi{10.1134/S000511790603012X}

\bibitem{CheLog10ARC}
Chebotarev, P.Y., Loginov, A.K., Tsodikova, Y.Y., Lezina, Z.M., Borzenko, V.I.:
  Analysis of collectivism and egoism phenomena within the context of social
  welfare. Automation and Remote Control  \textbf{71}(6),  1196--1207 (2010).
  \doi{10.1134/S0005117910060202}

\bibitem{CheMal18opt}
Chebotarev, P.Y., Malyshev, V.A., Tsodikova, Y.Y., Loginov, A.K., Lezina, Z.M.,
  Afonkin, V.A.: The optimal majority threshold as a function of the variation
  coefficient of the environment. Automation and Remote Control
  \textbf{79}(4),  725--736 (2018). \doi{10.1134/S0005117918040136}

\bibitem{CheTsoLog18}
Chebotarev, P.Y., Tsodikova, Y.Y., Loginov, A.K., Lezina, Z.M., Afonkin, V.A.,
  Malyshev, V.A.: Comparative efficiency of altruism and egoism as voting
  strategies in stochastic environment. Automation and Remote Control
  \textbf{79}(11),  2052--2072 (2018). \doi{10.1134/S0005117918110097}

\bibitem{EnelowiHinich84}
Enelow, J.M., Hinich, M.J.: The Spatial Theory of Voting: An Introduction.
  Cambridge University Press, Cambridge (1984)

\bibitem{Galasso02EconSecu}
Galasso, V., Profeta, P.: The political economy of social security: A survey.
  European Journal of Political Economy  \textbf{18}(1),  1--29 (2002).
  \doi{10.1016/S0176-2680(01)00066-0}

\bibitem{Kovalev10a}
Kovalev, S.A.: Civic responsibility of intellectuals ({Political} idealism and
  real politics: The challenge of the 21st century). Novaya Gazeta  (2010),
  \url{https://novayagazeta.ru/articles/2010/03/12/4341-grazhdanskaya-otvetstvennost-intellektualov}

\bibitem{Kranich01AltruIncome}
Kranich, L.: Altruism and the political economy of income taxation. Journal of
  Public Economic Theory  \textbf{3}(4),  455--469 (2001).
  \doi{10.1111/1097-3923.00078}

\bibitem{Levine98Modeling}
Levine, D.K.: Modeling altruism and spitefulness in experiments. Review of
  Economic Dynamics  \textbf{1}(3),  593--622 (1998).
  \doi{10.1006/redy.1998.0023}

\bibitem{Lindenberg01}
Lindenberg, S.: Social rationality versus rational egoism. In: Turner, J.H.
  (ed.) Handbook of Sociological Theory, chap.~29, pp. 635--668. Kluwer
  Academic/Plenum Publisher, New York (2001)

\bibitem{MaksChe20}
Maksimov, V.M., Chebotarev, P.Y.: Voting originated social dynamics: {Quartile}
  analysis of stochastic environment peculiarities. Automation and Remote
  Control  \textbf{81}(10),  1865--1883 (2020). \doi{10.1134/S0005117920100069}

\bibitem{Malyshev21optimal}
Malyshev, V.: Optimal majority threshold in a stochastic environment. Group
  Decision and Negotiation  \textbf{30}(2),  427--446 (2021).
  \doi{10.1007/s10726-020-09717-8}

\bibitem{Margolis84Selfishness}
Margolis, H.: Selfishness, Altruism, and Rationality. University of Chicago
  Press, Chicago (1984)

\bibitem{McKelvey76}
McKelvey, R.D.: Intransitivities in multidimensional voting models and some
  implications for agenda control. Journal of Economic Theory  \textbf{12}(3),
  472--482 (1976). \doi{10.1016/0022-0531(76)90040-5}

\bibitem{Mirkin79}
Mirkin, B.G.: {Group Choice}. V.H. Winston \& Sons, Washington D.C.
  (distributed by Halsted Press Division of John Wiley \& Sons, N.Y.) (1979)

\bibitem{PavlovaRigobon10}
Pavlova, A., Rigobon, R.: An asset-pricing view of external adjustment. Journal
  of International Economics  \textbf{80}(1),  144--156 (2010).
  \doi{10.1016/j.jinteco.2009.09.003}

\bibitem{Roberts77Voting}
Roberts, K.W.S.: Voting over income tax schedules. Journal of Public Economics
  \textbf{8}(3),  329--340 (1977). \doi{10.1016/0047-2727(77)90005-6}

\bibitem{Romer75IndWelf}
Romer, T.: Individual welfare, majority voting, and the properties of a linear
  income tax. Journal of Public Economics  \textbf{4}(2),  163--185 (1975).
  \doi{10.1016/0047-2727(75)90016-X}

\bibitem{TsChLo20Cambr}
Tsodikova, Y., Chebotarev, P., Loginov, A.: Modeling responsible elite. In:
  Aleskerov, F., Vasin, A. (eds.) Recent Advances of the Russian Operations
  Research Society, chap.~6, pp. 89--110. Cambridge Scholars Publishing,
  Newcastle upon Tyne (2020)

\bibitem{TsoChe24E}
Tsodikova, Y.Y., Chebotarev, P.Y.: Modeling society with a responsible elite.
  Journal of the New Economic Association (1 (66)),  12--35 (2025).
  \doi{10.31737/22212264_2025_1_12-35}, in Russian

\bibitem{Weron98}
Weron, A., Weron, R.: In\.zynieria Finansowa. Wydawnictwo Naukowo-Techniczne,
  Warszawa (1998)

\end{thebibliography}
\label{Lastpage}
\label{lastpage}
\end{document}